\documentclass[journal]{IEEEtran}
\usepackage{amsmath,amsfonts,amssymb}
\usepackage{array}
\usepackage{textcomp}
\usepackage{stfloats}
\usepackage{url}
\usepackage{verbatim}
\usepackage{graphicx}
\usepackage{float} 
\usepackage{subfigure}
\graphicspath{{/}}
\usepackage{multirow}
\usepackage{makecell}
\usepackage{hyperref}
\hypersetup{hidelinks}

\usepackage{adjustbox}  
\usepackage{calc}  

\usepackage{svg}
\usepackage[ruled,vlined,linesnumbered]{algorithm2e}
\SetAlFnt{\small}
\newcolumntype{C}[1]{>{\centering\arraybackslash}m{\dimexpr#1\textwidth-\tabcolsep}}
\renewcommand{\arraystretch}{2.5}
\definecolor{green}{RGB}{0, 150, 0}

\newtheorem{myDef}{Definition}

\usepackage{threeparttable}   
\usepackage{xcolor}

\begin{document}
\title{Population-Based Search Method Using Uncertainty-related Pareto Front for Robust Multi-objective Optimization}

\author{Lihong Xu~\IEEEmembership{} and Wenxiang Jiang~\IEEEmembership{}
	\thanks{This work was supported in part by the National Natural Science Foundation of China (Grant No. 62373286). ({\it Corresponding author: Lihong Xu)}}
	\thanks{Lihong Xu and Wenxiang Jiang are with the School of Electronic and Information Engineering, Tongji University, Shanghai 201804, China, Email: xulhk@163.com; wxjiangmail@163.com.}
	}

\maketitle

\begin{abstract}
Traditional robust multi-objective optimization methods typically prioritize convergence while treating robustness as a secondary consideration. This approach can yield solutions that are not genuinely robust optimal under noise-affected scenarios. Furthermore, compared to population-based search methods, determining the robust optimal solution by evaluating the robustness of a single convergence-optimal solution is also inefficient. To address these two limitations, we propose a novel Uncertainty-related Pareto Front (UPF) framework that balances robustness and convergence as equal priorities. Unlike traditional Pareto Front, the UPF explicitly accounts for decision variable with noise perturbation by quantifying their effects on both convergence guarantees and robustness preservation equally within a theoretically grounded and general framework.
Building upon UPF, we propose RMOEA-UPF—a population-based search robust multi-objective optimization algorithm. This method enables efficient search optimization by calculating and optimizing the UPF during the evolutionary process.  Experiments on nine benchmark problems and a real-world application demonstrate that RMOEA-UPF consistently delivers high-quality results. Our method's consistent top-ranking performance indicates a more general and reliable approach for solving complex, uncertain multi-objective optimization problems. Code is available at: \url{https://github.com/WenxiangJiang-me/RMOEA-UPF}.

\end{abstract}


\section{Introduction}\label{introduction}
\begin{figure}[!h]
	\centering
	\includegraphics[scale=0.35]{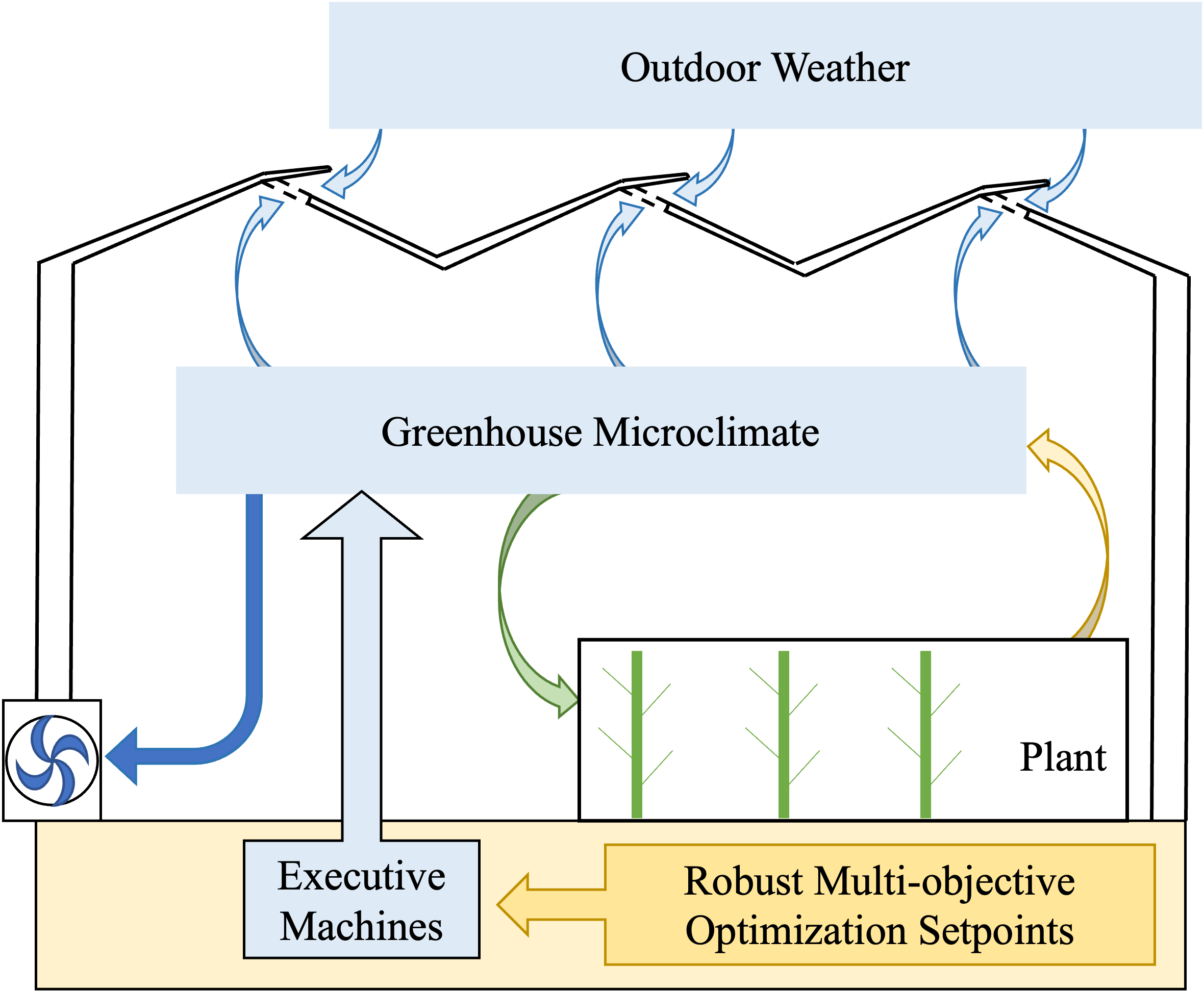}
	\caption{Greenhouse Microclimate Optimization Control System}
	\label{greenhouse}
	\vspace{-15pt}
\end{figure}

Over the years, multi-objective evolutionary algorithms (MOEAs) have demonstrated remarkable effectiveness in solving complex optimization challenges \cite{996017,zitzler2001spea2,4358754}. However, real-world optimization problems inherently involve uncertainties. Consider greenhouse microclimate optimization control system shown in Fig. \ref{greenhouse}, where we aim to maximize crop yield ($f_1$) while minimizing energy consumption ($f_2$) through optimal microclimate (temperature, $CO_2$ concentration, and light intensity) regulation. This represents a classic multi-objective optimization problem where objectives depend on microclimate states influenced by unpredictable weather patterns. As multi-month weather forecasts remain unreliable, these uncertainties create significant challenges for deriving robust Pareto optimal solutions – a critical gap in current robust multi-objective optimization research.
The challenge of uncertainty is not unique to agriculture; it is a prevalent issue across many critical engineering and economic domains. For instance, in modern energy systems, planners employ robust multi-objective optimization to manage demand response programs effectively, navigating the inherent volatility of renewable energy sources and market prices to balance retailer costs with consumer benefits \cite{khalili2023robust}. Similarly, in the information technology sector, the efficient operation of data center microgrids relies on robust optimization to handle the dual uncertainties of fluctuating wind power generation and unpredictable computational workloads, simultaneously minimizing costs and resource waste \cite{lian2023robust}. These examples underscore the critical need for advanced, general-purpose algorithms capable of finding truly robust solutions in complex, real-world scenarios.

\begin{figure}[!h]
	\centering
	\includegraphics[scale=0.7]{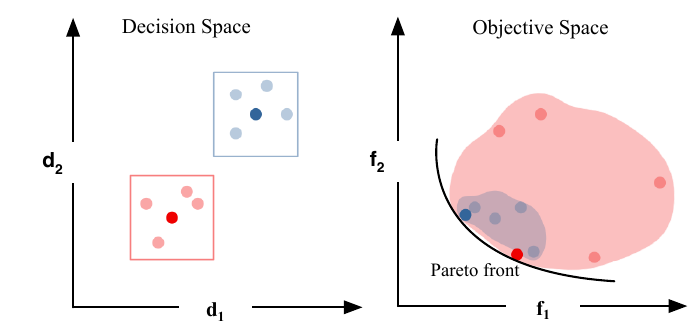}
	\caption{Diagram of traditional robust multi-objective optimization problems.}
	\label{deb robust}
	\vspace{-12pt}
\end{figure}
Current approaches to robust multi-objective optimization under input uncertainty primarily concentrate on robust optimization methods that evaluate the robustness of individual multi-objective optimization solutions. Fig. \ref{deb robust} illustrates the robust multi-objective optimization problem defined by traditional methods. Assuming the decision variables have two dimensions, denoted by $d_{1}$ and $d_{2}$, and the two objective functions are represented by $f_{1}$ and $f_{2}$, respectively. First, a solution is selected on the Pareto Front (favoring convergence). Taking $x_1$ (indicated by the point) and $x_2$ (indicated by the blue point) as an example, when faced with the same noise, the range of variation in the objective space for solution $x_2$ is smaller than that of solution $x_1$ (The red area represents the range of variation for solution $x_1$, while the blue area represents the range of variation for solution $x_2$.), leading to $x_2$ being selected as the robust optimal solution. The principal focus of this research lies in designing robustness evaluation metrics and developing methods to quantify robustness. The former involves designing robustness measures using statistical indicators (such as expectation and variance \cite{beyer2007robust,Das2000ROBUSTNESSOF}) or their combinations, including simple aggregations of expectation and variance \cite{10.1115/1.4001526,lei2013robust}, weighted sums of expectation and variance \cite{sun2011crashworthiness,1996Robust}, and ratios of expectation and variance \cite{2015Six} to assess the robustness of potential solutions. The latter concerns quantifying robustness and employs multiple sampling methods, such as Monte Carlo sampling \cite{Tomoyuki2002Monte}, to estimate a solution's robustness by averaging the fitness values of multiple neighboring samples. Given the high computational cost associated with multiple sampling methods, research focused on reducing the computational load of such methods is also progressing (e.g., using surrogate models or alternative sampling techniques \cite{5586235,Liu2016SurrogateAssistedUO}). 

However, these traditional methods face three issues: First, searching for robust solutions on the Pareto Front may overlook solutions that have strong robustness but slightly inferior convergence, resulting in poor diversity of solutions. This fundamental limitation restricts their ability to handle complex real-world problems requiring diverse solution sets and adaptive tradeoff analysis. Second, these methods guarantee robustness only through noise perturbation assessments on individual solutions, which is highly inefficient. Considering that traditional multi-objective optimization can employ non-dominated approaches for population-based search optimization, there is an urgent need to develop a class of population-based search optimization methods tailored for uncertain multi-objective optimization problems. Third, the fundamental idea of these methods prioritizes convergence while treating robustness as an ancillary preference for solution selection, which does not align with the essential definition of robust multi-objective optimization. Deb's Type I robustness \cite{6792442} calculates average objective values from multiple neighborhood samples, effectively prioritizing solutions with better mean convergence under perturbations. 

\begin{figure}
	\centering
	\includegraphics[scale=0.35]{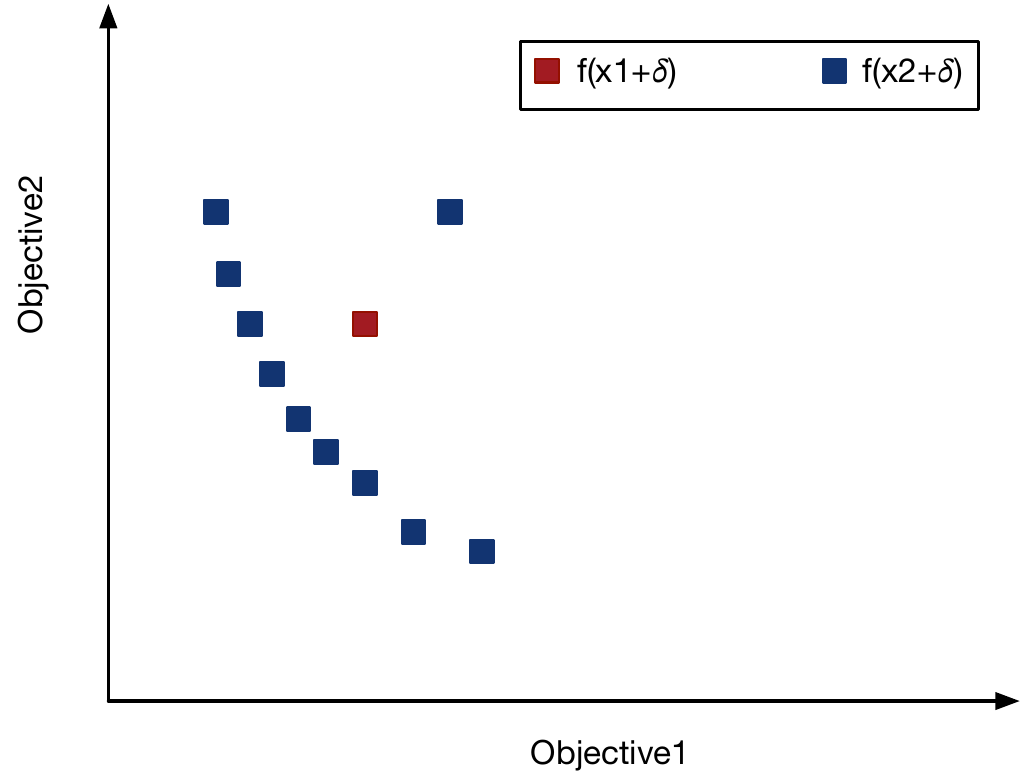}
	\caption{Diagram of robustness comparison}
	\label{solutionab}
	\vspace{-10pt}
\end{figure}

Consider two solutions under 10 noise disturbances as shown in Fig. \ref{solutionab}: We assume that solution \(x_1\) experiences ten noise perturbations, with a \(L^1\) norm value of 2 in the objective space for each perturbation, while solution \(x_2\) under the same conditions has nine \(L^1\) norm values of 1 and one \(L^1\) norm value of 5 in the objective space. Consequently, the average noise perturbation value for solution \(x_1\) is 2, while that for solution \(x_2\) is 1.4. According to the qualitative definition of robust multi-objective optimal solutions, a solution demonstrates better robustness if the variation in objective values is smaller under noise perturbations. Solution \(x_1\)'s consistent response to every perturbation indicates a stronger resistance to disturbances; hence, it can be considered more robust than solution \(x_2\). However, if we apply Type I robustness definitions, which consider the average value of solutions after noise perturbations, solution \(x_2\) has a lower average perturbation value under noise perturbations. Therefore, under this definition, \(x_2\) is considered the more robust solution, which conflicts with the definition of robust multi-objective optimization problems. The main reason for this conflict is that existing studies often prioritize convergence too highly even using convergence to assess robustness when exploring robust multi-objective optimization problems with noise perturbations in decision variables, treating robustness as an additional factor. This clearly contradicts the essential definition of robust multi-objective optimization problems. To effectively address robust multi-objective optimization issues, it is crucial to balance considerations of robustness and convergence; in other words, both should be regarded as equally important.

To overcome these fundamental issues, this study proposes the Uncertainty-related Pareto Front (UPF). The UPF framework marks a paradigm shift: instead of evaluating the robustness of individual solutions post-optimization, it redefines the core task as directly optimizing a non-dominated front that inherently embeds robustness guarantees for the entire population. This conceptual framework extends robust multi-objective optimization from single-preference analysis to comprehensive solution set optimization. Building upon UPF, we develop a population-based search algorithm that leverages historical performance data. Our approach presents three key innovations:
\begin{enumerate}
	\item We conduct a thorough analysis of current research on robust multi-objective optimization, highlighting significant misconceptions regarding the equal consideration of robustness and convergence, as well as the limitations of inefficient searches that focus on the robustness preference of a single solution.
	\item We propose the Uncertainty-related Pareto Front (UPF) concept, a theoretically grounded framework that uniquely enables the co-equal treatment of convergence and robustness. This provides a principled foundation for designing advanced algorithms, particularly population-based search methods, for robust multi-objective optimization.
	\item We propose RMOEA-UPF, a novel population-based algorithm specifically designed to optimize the UPF. It features an innovative archive-centric framework where the elite archive acts as the core population. Unlike traditional methods that maintain separate populations and archives, our approach generates parents directly from this elite archive, tightly integrating the selection of high-performing solutions with the creation of new candidates. This mechanism ensures an efficient and direct search for solutions that are superior in both convergence and robustness.
\end{enumerate}

The remainder of this paper is organized as follows: Section \ref{relatedwork} surveys existing approaches to robust multi-objective optimization. Section \ref{idea} formally introduces our core concept, the Uncertainty-related Pareto Front (UPF). Section \ref{method} details the proposed RMOEA-UPF, an archive-centric algorithm specifically designed to efficiently optimize the UPF. Section \ref{experiments} empirically validates the superiority of our algorithm through comprehensive experiments on benchmark problems and a real-world application. Finally, Section \ref{conclusion} concludes the paper by summarizing our contributions and discussing broader implications and future research directions.

\section{Related Work}\label{relatedwork}
When considering multi-objective optimization problems, various conflicting objectives may co-exist. The formulation of such problems is presented in Eq. (\ref{mop}), considering the minimization problem without loss of generality.
\begin{equation}
	\label{mop}
	\begin{split}
		{\min}\quad & {F(x)}=\left ( f_{1}({x}),f_{2}({x}),...,f_{M}({x})  \right )\\
		s.t. \quad&{x}\in \Omega
	\end{split}  
\end{equation}
where ${x}=(x_{1}, x_{2}, ..., x_{D})$ is a $D$-dimensional solution, $M$ is the number of objectives, and $\Omega\subseteq {R}^{n}$ represents the decision search space. 

In this article, we primarily focus on scenarios characterized by uncertainty in decision variables. The form of the Robust Multi-objective Optimization Problem (RMOP) with noisy decision variables is defined as
\begin{equation}
	\label{rmop}
	\begin{split}
		{\min}\quad & {F(x+\delta)}=( f_{1}(x+\delta),f_{2}(x+\delta),...,f_{M}(x+\delta) )\\
		s.t. \quad&\{x+\delta\}\in \Omega
	\end{split} 
\end{equation}
where ${\delta}=(\delta_{1}, \delta_{2}, ..., \delta_{D})$ is a $D$-dimensional noise vector. Given the maximum disturbance degree $\delta^{\max}=(\delta_{1}^{\max}, \dots, \delta_{n}^{\max})$, there exists $-\delta_{i}^{\max}\le \delta_{i}\le\delta_{i}^{\max}$ where $i\in \{1,\dots,D\}$.

In robust optimization, the primary goal is to ensure solutions meet robustness criteria while enhancing their convergence. Compared to traditional evolutionary optimization, robust multi-objective evolutionary algorithms (RMOEAs) better address practical problems with uncertainties. Current research focuses on comparing solution robustness and Pareto dominance under uncertainty, involving robustness metrics design and quantification methods, which fall into three main categories.

The first category uses expectation and variance metrics, requiring extensive function evaluations to estimate solution performance via neighboring fitness values. For example, expected hypervolume improvement (EHVI) \cite{2006Single} maximizes hypervolume under noise, while qNEHVI \cite{2021Parallel} integrates prior evaluation uncertainty. Methods like Monte Carlo integration \cite{1665030,738986,1438403,tsutsui1997genetic} replace original objectives with integrated performance-expectation functions. Researchers also use weighted averages based on disturbance probabilities \cite{branke1998creating} or variance metrics to measure objective deviation \cite{gaspar2008robustness,jin2003trade}. Type I robust optimization \cite{6792442} computes mean effective objectives from neighboring solutions. Combinations of statistical measures (mean, variance, standard deviation) \cite{6867354,lei2013robust,sun2011crashworthiness} and robust dominance concepts (e.g., dominance robustness, preference robustness \cite{5557781}, r-dominance \cite{li2015new}) further quantify robustness. The qNPAREGO method \cite{2020Differentiable} integrates quantile optimization for risk-aware decision-making, while Multi-Value-at-Risk (MVaR) \cite{2012Multivariate} translates probabilities into objective bounds but faces computational challenges. Ro-MOBO \cite{daulton2022robust} optimizes MVaR via Bayesian frameworks, though MVaR points may concentrate, reducing evaluation accuracy.

The second strategy focuses on constraint-oriented reliability assessment, evaluating robustness through geometric proximity to constraint boundaries. The Most Probable Point (MPP) methodology \cite{hasofer1974exact} and derivative techniques (e.g., fast performance measure approach \cite{gee2016solving}) use first/second-order approximations \cite{6867354,hohenbichler1987new}. Probabilistic evaluations across generations quantify compliance likelihood with design specs \cite{5491151,coelho2014probabilistic}, emphasizing constraint adherence and perturbation resilience.

The third strategy addresses the dual objectives of convergence and robustness through fitness function modification. This approach employs probabilistic dominance criteria \cite{coelho2014probabilistic,coelho2011multi} to rank solutions, complemented by robust dominance metrics derived from set coverage measures \cite{li2015new}. Central to this framework is the concept of dominance robustness, which quantifies a solution's capacity to maintain its Pareto Front positioning under perturbations \cite{5557781}. Within the Multi-objective Evolutionary Algorithm based on Decomposition (MOEA/D) framework \cite{4358754}, the Decomposition-Based Evolutionary Algorithm for Robust Optimization (DBEA-r) \cite{6867354} introduces a six-sigma quality measure that synthesizes expected values with standard deviations of feasibility and performance functions. A coevolutionary extension, the Coevolutionary Robust MOEA/D (C-RMOEA/D) \cite{7743846}, implements dual-population dynamics: one population optimizes decision variables while the other governs perturbation thresholds. This coevolutionary mechanism simultaneously evolves decision variables while enhancing robustness through perturbation thresholds governed by performance degradation analysis in subproblems. The RMOEA framework \cite{8419222} adopts a phased methodology, prioritizing exhaustive exploration of the objective space for optimality before systematically identifying robust regions to construct the final Pareto Front. Further advancing this framework, the Promising-Region-Based EMO Algorithm (PREA) \cite{yuan2020investigating} leverages ratio-based indicators to identify high-potential regions in the objective space. Candidate solutions within these regions undergo normal plane projection to eliminate suboptimal individuals, thereby refining population quality. The Robust multi-objective Evolutionary Algorithm with Robust Enhancement (MOEA-RE) \cite{he2019evolutionary} proposes an archive updating mechanism that synergistically integrates robustness and optimality with reduced computational overhead, culminating in a dedicated robust optimal front-building strategy. More recently, the LRMOEA algorithm \cite{10612838} tackles large-scale robust multi-objective problems by utilizing a dual-structure where the main population's evolution is driven exclusively by the pursuit of convergence. Concurrently, a separate archive is maintained to manage robustness by selecting and preserving solutions that perform well under uncertainty. While robust knowledge from the archive is used to assist the search process, the primary optimization pressure remains on convergence, effectively treating robustness as a parallel but separate objective.
\begin{figure*}[b]
	\centering
	
	\subfigure[UPF diagram (alpha=0.9)]{\includegraphics[width=0.47\linewidth,height=0.27\textheight]{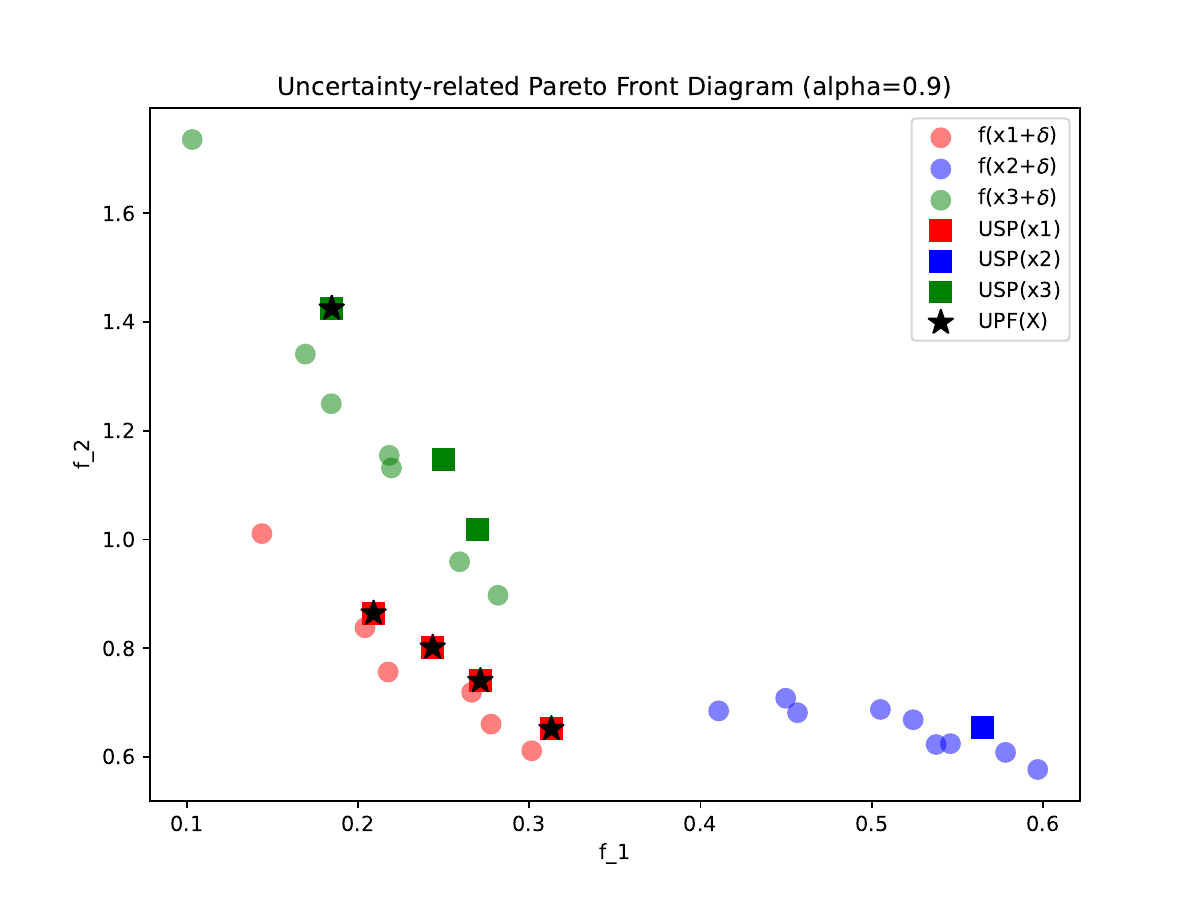}}
	\subfigure[UPF diagram (alpha=0.5)]{\includegraphics[width=0.47\linewidth,height=0.27\textheight]{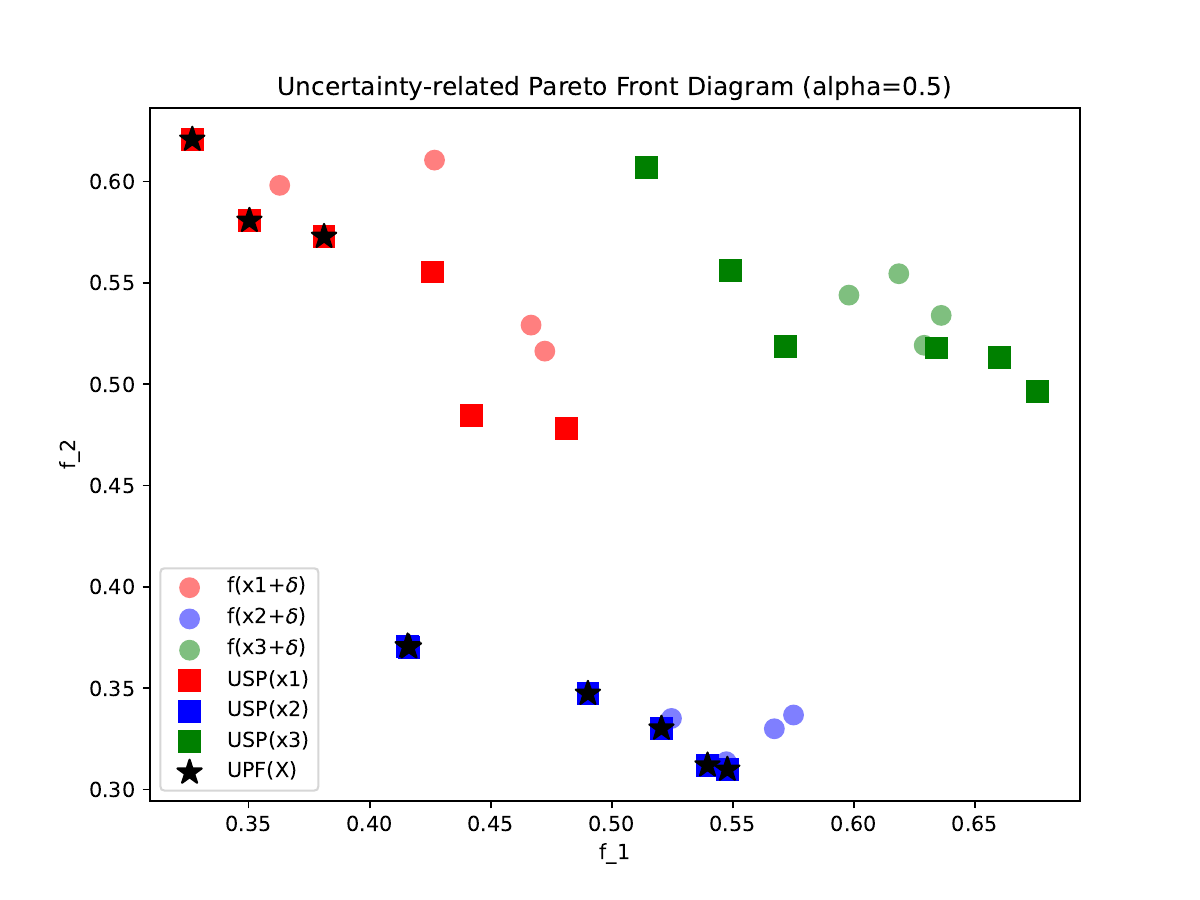}}
	\centering
	\caption{Diagram of UPF}
	\label{upf_diagram}
	\vspace{-15pt}
\end{figure*}

While the above three categories of methods demonstrate methodological diversity, they share a common limitation that hinders a truly holistic search. Although population-based search is a foundational paradigm in traditional multi-objective optimization, its direct application in the robust domain has been challenging. This is because many of the aforementioned methods treat convergence and robustness as separate, sequential steps. They typically focus first on finding solutions using convergence as the primary criterion, and only then evaluate the robustness of these single solutions as a secondary concern. This approach, which prioritizes convergence while treating robustness as a supplementary condition, inherently prevents a truly simultaneous search where both objectives are considered equally.

A separate category optimizes by analyzing decision variables. RMOEA-DVA \cite{liu2021decision} classifies variables into LDRVs (low dominance robustness) and HDRVs (high dominance robustness), optimizing them with original objectives or robustness metrics respectively. CNSDE/DVC \cite{8359103} categorizes high-dimensional variables into DV1 (critical for robustness) and DV2 (minimal impact), applying targeted strategies. However, these methods lack generality. RMOEA-SuR \cite{Jiang2025} uses survival rate (based on non-dominated status under noise) as an additional objective, but still relies on convergence to characterize robustness. Thus, a framework for simultaneous convergence-robustness optimization remains essential.

\section{Transform Robust Multi-objective Optimization into UPF Optimization}\label{idea}
Based on the preceding analysis, we contend that for devising an effective approach to robust multi-objective optimization, robustness and convergence must be treated as two core elements that demand consideration of equal importance. To address the limitations of methods that assess robustness as a secondary concern to convergence, we leverage probability as a direct measure of robustness. Our approach, therefore, seeks a solution set that is Pareto-optimal in terms of convergence, subject to a quantifiable, probabilistic robustness guarantee.

As illustrated in Fig. \ref{upf_diagram}, we take the TP1 test problem (selected from \cite{6792442}) for minimizing a bi-objective optimization problem as an example, assuming three decision vectors (or solutions) \( x_1, x_2, \) and \( x_3 \). We perform ten noise samplings on these three decision vectors (applying ten perturbations to the decision variables and then computing the corresponding objective space vectors, where the ten noise perturbations are identical for each decision variable). In the figure, the red, blue, and green solid circles represent the ten noise points of \( x_1, x_2, \) and \( x_3 \), respectively (there is a chance that some noise points are covered by solid squares, which will be detailed later). We then aim to identify the ``most conservative'' support point from each set of objective space vectors, representing the worst-case performance of the decision variable under noise perturbation while ensuring robustness. For each solution, we estimate a dominance probability for each of its noisy points. This is done by calculating the fraction of other noisy points from the same solution that it dominates. Letting the confidence level \(\alpha\) be 0.9, we select points with dominance probabilities less than or equal to \(1 - \alpha\) (i.e., 0.1), and among these points, find those whose dominance probability is closest to \(1 - \alpha\); these are defined as the ``most conservative'' support points. Because under noise perturbations, the performance of the decision variable in the objective space is better than or equal to these points with probability at least \(\alpha\), we term these points Uncertain \(\alpha\)-Support Points (USP). In the figure, the solid squares in red, blue, and green correspond to USP(\(x_1\)), USP(\(x_2\)), and USP(\(x_3\)), respectively. These USP represent the ``worst-case'' performance of the corresponding decision variables under noise perturbation that satisfy the robustness condition; that is, the decision variables can achieve performance no worse than these USP with probability at least \(\alpha\) under noise perturbation. Hence, USP simultaneously guarantee robustness and convergence. For example, in Fig. \ref{upf_diagram}(a), we can find that  decision variables \(x_1, x_2,\) and \(x_3\) have four, one, and three points in USP, respectively.

\begin{myDef}\label{usp}
	For a decision variable $x$ and a confidence level $\alpha \in [0,1]$, the set of the following non-dominated points in the objective space image $F(x+\delta)$ under uncertain perturbation is given by
	\begin{equation*}
		\mathcal{F} = 
		\left\{ 
		z \in \mathbb{R}^M \ 
		\middle| \ 
		P\bigl[ z \prec f(x+\delta) \bigr] \le (1-\alpha) 
		\right\}  \nonumber
	\end{equation*}
	The Uncertain \(\alpha\)-Support Points (USP) for a solution $x$ and confidence level  is defined as the point closest to the probability $ (1-\alpha) $ within $ \mathcal{F}$: 
	
	\begin{multline}
		\mathrm{USP}(x, \alpha) = \left\{ z^* \in \mathbb{R}^M \,\bigg|\, \right. \\
		z^* = \mathrm{arg}\min_{z \in \ \mathcal{F}} 
		\left. \Bigl\lvert \mathbb{P}\bigl[ z \prec f(x+\delta) \bigr] - (1-\alpha) \Bigr \rvert \right\} \nonumber
	\end{multline}
\end{myDef}
Here, \(\prec\) denotes the dominance relation; for example, \(\alpha \prec b\) indicates that \(a\) dominates \(b\). The symbol $|\cdot|$ represents the absolute value. The parameter \(\alpha\) is the robustness confidence level set according to user preference. The USP serves as a robust performance benchmark for a solution x. It represents the objective vector for which one can be $\alpha$-confident (e.g., $alpha$=0.9) that the solution's actual performance under noise will be better than or equal to this benchmark. It effectively defines a ``worst-case" scenario under a given confidence level.

According to the Definition \ref{usp}, USP are the vectors in the objective space closest to the probability \(1 - \alpha\), such that when the decision variable faces noise perturbation \(\delta\), the convergence performance in the objective space is not worse than the vector \(\text{USP}(x, \alpha)\) with a probability greater than or equal to \(\alpha\). We can apply the same noise perturbation to a set of decision variables \(X\) and subsequently compute their \(\alpha\)-support points, denoted as \(\text{USP}(X, \alpha)\). These USP points represent the ``most conservative'' performance of the decision variables under noise perturbations, ensuring that, with at least a confidence level \(\alpha\), their convergence performance is no worse than these USP. Furthermore, the non-dominated set formed by these USP illustrates the population's best noise-resilience performance under noise perturbations. Based on the survival-of-the-fittest principle in optimization algorithms, we can identify the optimal solutions that balance robustness and convergence by finding the non-dominated set of the USP collection. Therefore, this non-dominated set of the USP collection holds a significance analogous to the traditional Pareto Front in multi-objective optimization, and we define it as the Uncertainty-related Pareto Front (UPF) of the following uncertain problem, shown as black solid pentagrams in Fig. \ref{upf_diagram} (the noise generation method is Monte Carlo sampling, with perturbations as uniform random noise, and the perturbation radius for each component of the decision variable set to 10\% of their respective domain widths). The definition of UPF is as follows: 
\begin{myDef}\label{upf}
	The non-dominated set of all USP of the decision variable set \(X\) under perturbation is called the Uncertainty-related Pareto Front of \(X\) at confidence level \(\alpha\):
\begin{equation}
\mathrm{UPF}(X,\alpha) = \left\{ \mathrm{USP}(x,\alpha) \;\middle|\; 
\substack{
	\not \exists x^{'} \in X, \\ 
	\mathrm{USP}(x^{'},\alpha) \prec \mathrm{USP}(x,\alpha)
} \right\} 
\nonumber
\end{equation}
\end{myDef}

Each USP is a robust benchmark for a single solution, therefore, the UPF is the Pareto-optimal front of these benchmarks for the entire population. A point on the UPF represents a state where no other solution in the population can offer a better robust benchmark (i.e., be better on at least one objective without being worse on any other). The UPF thus represents the best trade-offs between objectives achievable by the population under the specified robustness guarantee $\alpha$. In Fig. \ref{upf_diagram}(a), we can find that there are four USP of \(x_1\) and one USP of \(x_3\) located on the UPF, while the USP of \(x_2\) lies off the UPF. To advance to the next generation, we would prioritize \(x_1\) and \(x_3\). In the final solution selection step, if the population size per generation is fixed and including both \(x_1\) and \(x_3\) exceeds this limit, we prioritize \(x_1\) over \(x_3\) because \(x_1\) has four USP located on the UPF, which is greater than the one associated with \(x_3\). For further illustration, Fig. \ref{upf_diagram}(b) shows the resulting USP and UPF when a less confidence level of $alpha$=0.5 is used, helping to clarify the role of $alpha$ in this framework. It is evident that robust multi-objective optimization problems involving input noise can be transformed into the task of searching for and optimizing the UPF. Specifically, through an evolutionary iterative process, the population is continuously updated—with the core goal of driving the population’s UPF toward the Pareto optimal direction—while persistently seeking a non-dominated solution set that balances robustness and convergence. This process ultimately leads to the acquisition of the optimal robust solution set.

The greatest significance of the UPF lies in the fact that, analogous to traditional multi-objective optimization algorithms, it allows us to convert robust multi-objective optimization problems affected by noise perturbations into a population-based search and optimization problem centered on the UPF. Through this optimization process, the approach enables the population to be optimized in a way that simultaneously accounts for both robustness and convergence, thereby ultimately yielding robust optimal solutions.

In this section, building on the preceding analysis of robust multi-objective optimization challenges under uncertainty, we innovatively propose the concept of the Uncertainty-related Pareto Front (UPF) tailored for uncertain problems. Specifically, the UPF denotes a set of solutions in the objective space where—after being subjected to noise perturbations—no alternative solution can achieve dominance under a predefined confidence level. Furthermore, by continuously calculating and optimizing the UPF throughout the optimization process, we can effectively implement population-based search optimization for robust multi-objective problems—a foundational design principle that will underpin the algorithm we present in the next section. 
\section{Proposed Algorithm}\label{method}
In this section, we propose a robust multi-objective evolutionary algorithm based on the UPF, hereinafter referred to as RMOEA-UPF. The core of RMOEA-UPF is a dynamic, fixed-size elite archive that not only preserves the best-performing solutions found so far but also actively drives the evolutionary search process. By iteratively refining this archive through a rigorous UPF evaluation of solution robustness and convergence, this approach enables the algorithm to simultaneously and efficiently optimize both the convergence and robustness of decision variables.

\subsection{Main Framework of RMOEA-UPF}
\begin{figure*}
	\centering
	\includegraphics[scale=0.6]{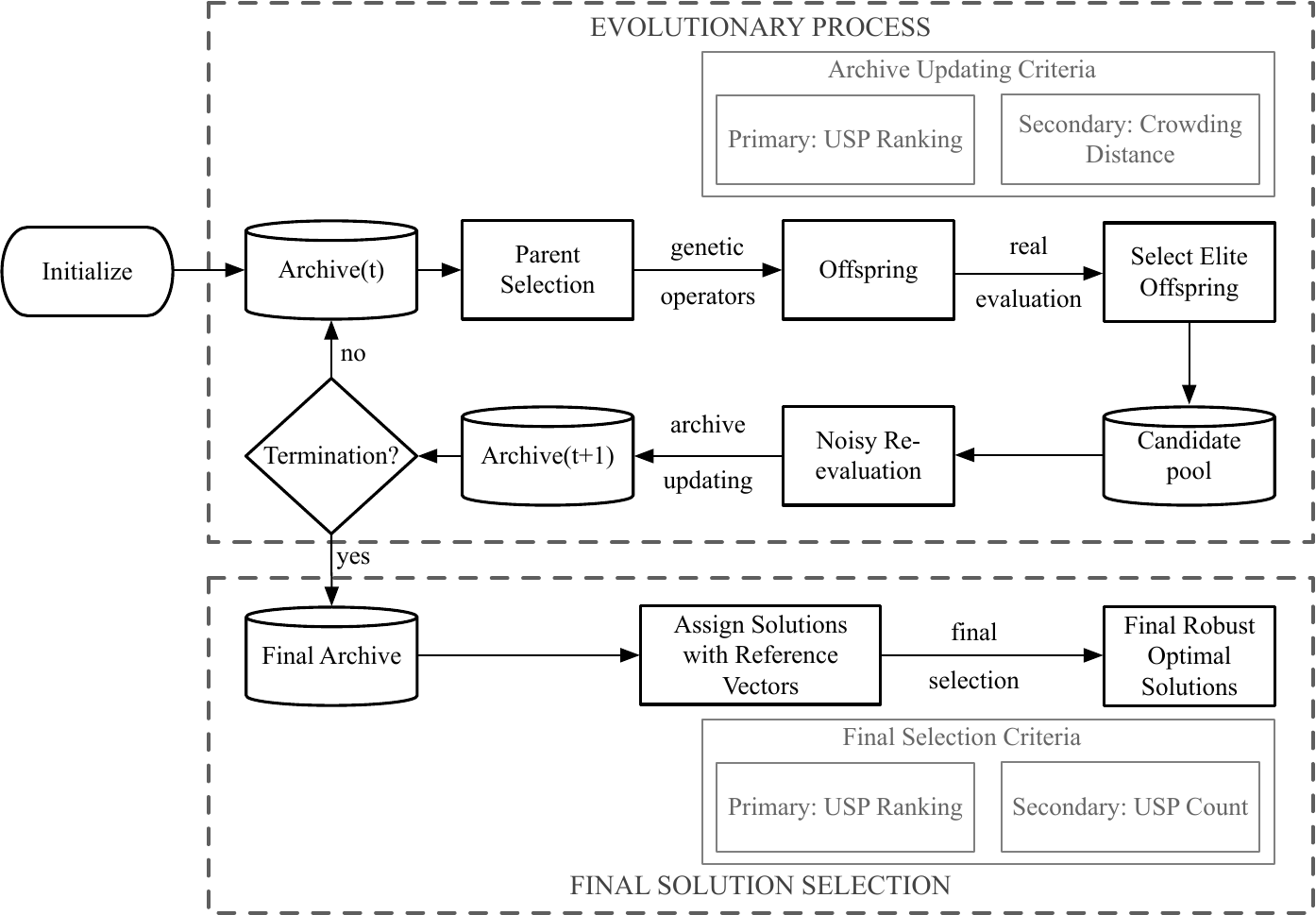}
	\caption{Main framework of RMOEA-UPF}
	\label{framework}
		\vspace{-15pt}
\end{figure*}

\begin{algorithm}[] 
	
	\SetKwData{Archive}{Archive} \SetKwData{P}{P} \SetKwData{O}{O} \SetKwData{Oelite}{O\_{elite}} \SetKwData{Pfinal}{P\_{final}} \SetKwFunction{UpdateArchive}{UpdateArchive} \SetKwFunction{FinalSelect}{FinalSolutionSelection} \SetKwInOut{Input}{Input}\SetKwInOut{Output}{Output}
	
	\Input{$MaxEvals$ (Maximum number of function evaluations), $N_{arc}$ (Archive size), $N_{pop}$ (Population size), $N_e$ (Elite offspring size), $\alpha$ (Confidence Level)}
	\Output{\Pfinal (Final robust optimal solution set)}
	\BlankLine
	
	$t \leftarrow 0$\tcp*{Initialize evaluation counter}
	\Archive, $evals\_used \leftarrow$ InitializeArchive{$N_{arc}$}\;
	$t \leftarrow t + evals\_used$\;
	
	\While{$t < MaxEvals$}{
		\P $\leftarrow$ SelectParents(\Archive, $N_{pop}$)\;
		\O $\leftarrow$ GenerateOffspring(\P)\;
		\O, $evals\_used \leftarrow$ EvaluateOffspring(\O)\;
		$t \leftarrow t + evals\_used$\;
		\Oelite $\leftarrow$ SelectEliteOffspring(\O, $N_e$)\;
		\If{$t < MaxEvals$}{
			\Archive, $evals\_used \leftarrow$ UpdateArchive(\Archive, \Oelite, $N_{arc}$, $\alpha$)\;
			$t \leftarrow t + evals\_used$\;
		}
	}
	
	\Pfinal $\leftarrow$ FinalSolutionSelection(\Archive)\;
	\Return{\Pfinal}\;
	\caption{Main Framework of RMOEA-UPF}
	\label{alg:main_framework}
	
\end{algorithm}

The main framework of RMOEA-UPF, depicted in Fig. \ref{framework}, operates through a single, archive-centric loop. This structure moves away from the traditional model of separate populations and archives. Instead, it employs an archive-centric approach where the main population is generated directly from the elite members of the archive. The algorithm's progression is then driven by a comprehensive, evaluation-based environmental selection that determines which of these new candidates will persist and update the archive. This ensures a direct and continuous lineage of high-quality, robust solutions throughout the optimization process.
\begin{algorithm*}[!b]
	\SetKwData{ArchivePrev}{Archive\_{(t)}}
	\SetKwData{Oelite}{O\_{elite}}
	\SetKwData{Pool}{CandidatePool}
	\SetKwData{ArchiveNew}{Archive\_{(t+1)}}
	\SetKwFunction{RealEval}{RealEvaluate}
	\SetKwFunction{UpdateHistory}{UpdateHistory}
	\SetKwFunction{CalculateUSP}{CalculateUSP}
	\SetKwFunction{NonDomSort}{NonDominatedSort}
	\SetKwFunction{GetRank}{GetFrontIndex}
	\SetKwFunction{CrowdingDistance}{CrowdingDistanceAssignment}
	\SetKwInOut{Input}{Input}\SetKwInOut{Output}{Output}
	
	\Input{\ArchivePrev (Archive of the previous generation), \Oelite (Elite offspring candidates), $N_{\text{arc}}$ (The maximum number of solutions in the archive), $\alpha$ (Confidence level)}  
	\Output{\ArchiveNew (New generation archive), $evals\_used$ (Number of evaluations used)}
	\BlankLine
	\tcp{Let a solution $s$ be a structure with attributes: $s.dec$, $s.history$, $s.USP$, $s.rank$}
	\Pool $\leftarrow$ \ArchivePrev $\cup$ \Oelite\;
	$evals\_used \leftarrow 0$\;
	\ForEach{solution $s$ in \Pool}{
		$obj_{\text{new}} \leftarrow$ \RealEval{$s.dec + \text{noise}$}\;  
		\UpdateHistory{$s, obj_{\text{new}}$}\;  
		$evals\_used \leftarrow evals\_used + 1$\;
	}
	
	\tcp{Step 1: Calculate USP and best rank for each solution}
	$AllUSP \leftarrow \bigcup_{s \in \Pool} $\CalculateUSP$(s.history, \alpha)$\;
	$USP_{\text{fronts}} \leftarrow$ \NonDomSort{$AllUSP$}\;  
	\ForEach{solution $s$ in \Pool}{
		$best\_rank \leftarrow \infty$\;  
		\ForEach{usp in $s.USP$}{
			$current\_rank \leftarrow$ \GetRank{$usp, USP_{\text{fronts}}$}\;  
			\If{$current\_rank < best\_rank$}{
				$best\_rank \leftarrow current\_rank$\;
			}
		}
	}
	\tcp{Step 2: Calculate crowding distance within each rank}
	$RankGroups \leftarrow \text{GroupBy}(\Pool, \text{by } s.best\_rank)$\;
	\ForEach{group in $RankGroups$}{
		$Objs \leftarrow \{s.obj \text{ for } s \in group\}$\;
		$Distances \leftarrow \CrowdingDistance{Objs}$\;
		\For{$i \leftarrow 1$ \KwTo $|group|$}{
			$group[i].crowding\_dist \leftarrow Distances[i]$\;
		}
	}
	\tcp{Step 3: Sort the pool and update archive}
	$SortedPool \leftarrow \text{Sort}(\Pool, \text{by } s.best\_rank \text{ then by } -s.crowding\_dist)$\;
	\ArchiveNew $\leftarrow SortedPool[1:N_{\text{arc}}]$\;
	
	\Return{\ArchiveNew, $evals\_used$}\;
	\caption{UpdateArchive}
	\label{alg:update_archive}
	
\end{algorithm*}

The algorithmic procedure is formalized into the following distinct steps, with the overall process outlined in Algorithm \ref{alg:main_framework}:

\begin{enumerate}
	\item \textbf{Initialization}: An initial archive is formed by generating a set of random solutions within the decision space. Each of these solutions undergoes function evaluation to obtain its initial objective vector. This collection of evaluated solutions constitutes the first-generation elite archive, and their evaluation results initialize their respective performance histories (Algorithm \ref{alg:main_framework}, line 2).
	
	\item \textbf{Parent Selection and Offspring Generation}: At the beginning of each generation, parent solutions are selected directly from the current elite archive (line 5). The top-ranked solutions in the archive are chosen to form the mating pool, ensuring that subsequent offspring are generated from individuals with proven high convergence and robustness. Then the selected parents are subjected to standard genetic operators to produce a new population of offspring solutions (line 6). Each new offspring solution is evaluated using objective functions (line 7). 
	
	\item \textbf{Archive Updating}: A new elite archive for the subsequent generation is constructed through a competitive environmental selection process (line 9). The newly evaluated offspring and the existing archive members are combined into a candidate pool. All unique solutions in this pool undergo an additional function evaluation with a single injection of random noise (within the predefined noise range) to enrich their performance history. Based on their complete historical data, their USP are calculated and used to rank them. The top-ranked solutions are then selected to form the next generation's archive, maintaining its fixed size (lines 10-12).
	
	\item \textbf{Final Solution Selection}: Once the algorithm's termination criterion (e.g., the maximum number of real evaluations) is met, a final selection mechanism is applied to the solutions in the terminal elite archive. This mechanism utilizes the comprehensive historical evaluation data of each solution, alongside reference vectors as auxiliary guidance, to construct the final robust optimal solution set, which is well-distributed and of a predefined size (lines 13-14).
\end{enumerate}

\subsection{Archive Updating Method}
The archive updating mechanism, detailed in Algorithm \ref{alg:update_archive}, is the core of RMOEA-UPF. It meticulously curates the elite set of solutions that guides the evolutionary search. This process is executed at the end of each generation and is designed to rigorously assess and select solutions based on their long-term demonstrated convergence and robustness. The procedure unfolds in a multi-stage sequence:

First, a preliminary screening of the newly generated offspring is conducted to identify high-potential individuals. Each new offspring solution is subjected to a single, deterministic real function evaluation without noise. The resulting objective vector is used to rank all offspring via non-dominated sorting. Based on this ranking, a fixed number of the best-performing offspring are selected as ``elite offspring candidates'' into the archive. A temporary candidate pool is formed by combining these elite offspring candidates with all the existing members of the current archive  (Algorithm \ref{alg:update_archive}, line 1).

Second, to obtain a more accurate assessment of both convergence and robustness, every solution within this candidate pool is subjected to one additional real function evaluation under a new noise instance. This step is critical as it progressively builds a rich performance history for each solution  (lines 3-6). For long-standing archive members, this history becomes a comprehensive record of their behavior across various noise conditions. For new candidates, this evaluation enriches their nascent performance history, which is crucial for subsequent, more accurate USP calculations.

Finally, the selection for the new archive is performed. For every solution in the candidate pool, its true Uncertain $\alpha$-Support Point (USP) set, $s.USP$, is calculated using its complete, updated history of real evaluations (line 7). A hierarchical ranking is then applied to all solutions (line 8). The primary ranking criterion is the non-domination level of a solution's USP when compared globally against all other USP in the pool (lines 9-14). A secondary criterion, the crowding distance, is then used as a tie-breaker for solutions within the same USP rank (lines 15-20). The crowding distance is calculated based on the solutions' deterministic objective values to promote diversity in the objective space. The archive for the next generation is then populated by selecting the top-ranked solutions based on this lexicographical comparison of USP rank and crowding distance, ensuring the survival of individuals that are not only robust but also contribute to a well-distributed front (lines 21-23).

\subsection{Final Robust Optimal Solution Set Construction}

\begin{algorithm}[]
	\SetKwData{ArchiveFinal}{Archive\_{final}}
	\SetKwData{Pfinal}{P\_{final}}
	\SetKwData{RefVecs}{RefVecs}
	\SetKwData{Niches}{Niches}
	\SetKwData{Unselected}{Unselected}
	\SetKwFunction{CalculateUSP}{CalculateUSP}
	\SetKwFunction{NonDomSort}{NonDominatedSort}
	\SetKwFunction{GetRank}{GetFrontIndex}
	\SetKwFunction{CountInFront}{CountInFront}
	\SetKwFunction{GenerateRefVecs}{GenerateReferenceVectors}
	\SetKwFunction{Associate}{AssociateToNiches}
	\SetKwFunction{SelectBestInNiche}{SelectBestInNiche}
	\SetKwInOut{Input}{Input}\SetKwInOut{Output}{Output}
	
	\Input{\ArchiveFinal (The final archive after iteration termination), $k$ (final set size), $\alpha$ (Confidence level)}
	\Output{\Pfinal (Final robust optimal solution set)}
	\BlankLine
	
	\tcp{Step 1: Calculate hierarchical rank for each solution}
	$AllUSP \leftarrow \bigcup_{s \in \ArchiveFinal} $\CalculateUSP$(s.history, \alpha)$\;
	$USP_{\text{fronts}} \leftarrow$ \NonDomSort{$AllUSP$}\;
	\ForEach{solution $s$ in $\ArchiveFinal$}{
		$best\_rank \leftarrow \min_{usp \in s.USP} \GetRank(usp, USP_{\text{fronts}})$\;
		$count \leftarrow \CountInFront(s.USP, USP_{\text{fronts}}[best\_rank])$\;
		$s.rank \leftarrow (best\_rank, -count)$\;
	}
	\tcp{Step 2: Niche-based selection for diversity}
	\RefVecs $\leftarrow$ \GenerateRefVecs{$k$}\;
	\Niches $\leftarrow$ \Associate{\ArchiveFinal, \RefVecs}\;
	\Pfinal $\leftarrow \emptyset$\;
	\ForEach{niche in \Niches}{
		\If{niche is not empty}{
			$s_{best} \leftarrow$ \SelectBestInNiche{niche, by $s.rank$}\;
			\If{$s_{best} \notin \Pfinal$}{
				\Pfinal $\leftarrow \Pfinal \cup \{s_{best}\}$\;
			}
		}
	}
	\tcp{Step 3: Fill remaining slots if necessary}
	\If{$|\Pfinal| < k$}{
		\Unselected $\leftarrow \ArchiveFinal \setminus \Pfinal$\;
		Sort \Unselected by $s.rank$\;
		$num\_to\_add \leftarrow k - |\Pfinal|$\;
		\Pfinal $\leftarrow \Pfinal \cup \Unselected[1:num\_to\_add]$\;
	}
	
	\Return{\Pfinal}\;
	\caption{FinalSolutionSelection}
	\label{alg:final_selection}
	
\end{algorithm}
Upon termination of the evolutionary process, a final selection procedure, detailed in Algorithm \ref{alg:final_selection}, is employed to distill a solution set of a desired size, k, from the terminal elite archive. This method is designed to yield a set of solutions that is not only of high quality in terms of robustness and convergence but also well-diversified across the objective space. This is accomplished by leveraging the rich historical data accumulated for each solution.

The procedure begins by calculating a USP set (Definition \ref{usp}) for each solution in the final archive from its complete performance history. This provides the most reliable possible characterization of its performance envelope based on all available information. A hierarchical rank is then assigned to each solution based on the non-domination level of its USP and the count of its USP on its best front, identical to the ranking metric used in the archive update (Algorithm \ref{alg:final_selection}, lines 1-6).

To ensure diversity, a set of k uniformly distributed reference vectors is generated in the objective space. Each solution from the archive is then associated with its nearest reference vector, determined by cosine distance. This partitions the solutions into k niches (lines 7-8). For each niche, the single associated solution with the best hierarchical rank is selected for the final set; if multiple solutions share the same best rank, the one with a higher count of USP on that front is chosen as a tie-breaker. This initial pass prioritizes filling each niche with its highest-quality candidate (lines 9-14).

If this initial selection results in fewer than k unique solutions (as some niches may have been empty or their best solution was already selected for another niche), the remaining slots are filled. The unselected solutions from the archive are sorted globally by their hierarchical rank (again using the USP count as a tie-breaker). The top-ranked unselected solutions are then iteratively added to the final set until its size reaches k. This two-phase process guarantees that the final output is a well-distributed set of the most convergent and robust solutions discovered by the algorithm (lines 15-20).

 \section{Experiments}\label{experiments}
In this study, we evaluate the performance of robust multi-objective optimization algorithms using the Uncertainty-related Pareto Front (UPF) on a suite of test problems. The UPF, defined as the non-dominated set of Uncertain $\alpha$-Support Points (USP), serves as a robust performance metric that quantifies an algorithm's optimal convergence under noise perturbations at a specified confidence level. By comparing the UPF generated by different algorithms, we can effectively assess their respective robustness and convergence capabilities. An algorithm is considered superior if its UPF is closer to the global optimum, indicating better overall performance.  Before the experiment begins, all algorithms are evaluated on the same noisy data to generate their respective UPF: $\mathrm{UPF}(\mathcal{X}_{k},\alpha)$ ($\mathcal{X}_{k}$ denotes the solution set of the k-th algorithm). Second, we construct a ``global UPF'' ($\mathrm{UPF_{global}}$) by taking the non-dominated set of all  $\mathrm{UPF}(\mathcal{X}_{k},\alpha)$ from all algorithms combined. This $\mathrm{UPF_{global}}$ serves as a comprehensive reference, representing the best robustness and convergence achieved across all algorithms under noise perturbation. 
	
To address the challenge that traditional performance metrics like Inverted Generational Distance (IGD) require a known UPF (for existing test problems with noise-perturbed decision variables, a known UPF is often unavailable.), we compute IGD by measuring the distance from ``global UPF'' ($\mathrm{UPF_{global}}$) to each algorithm's UPF ($\mathrm{UPF}(\mathcal{X}_{k},\alpha)$). A smaller IGD value indicates that an algorithm's UPF is more aligned with the $\mathrm{UPF_{global}}$, thus demonstrating superior robustness and convergence performance.

To further compare the performance of various algorithms, we modify the Generational Distance (GD) metric \cite{Lamont1999}. We propose the modified Generational Distance (mGD) to measure the average shortest distance from an algorithm's entire set of Uncertain \(\alpha\)-Support Points (USP) to the global Uncertainty-related Pareto Front ($\mathrm{UPF_{global}}$). The calculation process is as follows: first, the same noise data are added to the populations obtained from all algorithms, and evaluations using the real function are performed to derive each algorithm's USP set, denoted as $\mathrm{USP}(\mathcal{X}_{k},\alpha)$ (where $\mathcal{X}_{k}$ represents the solution set of the k-th algorithm). Then, we compute the average of the shortest distances from its $\mathrm{USP}(\mathcal{X}_{k},\alpha)$ to $\mathrm{UPF_{global}}$. Notably, mGD differs from the Inverted Generational Distance (IGD), which calculates the average of the shortest distances from points on $\mathrm{UPF_{global}}$ to the algorithm's UPF. Since the USP generated by a solution set are not perfectly non-dominated (i.e., some may be dominated by others within the set), mGD provides a more holistic assessment of an algorithm's robustness and convergence capabilities compared to IGD.

\begin{multline}\label{calculate-gd}
	\mathrm{mGD}\big( \mathrm{USP}(\mathcal{X}_{k},\alpha), \mathrm{UPF_{global}} \big) \\
	= \frac{1}{ |z_{k} | } 
	\sum_{z\in \mathrm{USP}(\mathcal{X}_{k},\alpha)} 
	\min_{u\in \mathrm{UPF_{global}}} \left \| z-u \right \|
\end{multline}
where $ | z_{k} |$ represents the number of points in  $\mathrm{USP}(\mathcal{X}_{k},\alpha)$.

\subsection{Experiment Settings}\label{settings}
Five typical comparison algorithms for robust optimization with noisy inputs were selected: LRMOEA (\cite{10612838}), MOEA-RE (\cite{he2019evolutionary}), RMOEA-SuR(\cite{Jiang2025}), NSGA2-DT1 (\cite{6792442}) and NSGA2 (\cite{996017}). With respect to the test functions, nine bi-objective test problems, TP1-TP9, were selected from Deb and Gupta \cite{6792442}, \cite{gaspar2014evolutionary}, each demonstrating various robust and optimal characteristics. 

In this comparative study, we evaluated the performance of all algorithms using the Uncertainty-related Pareto Front (UPF) with a confidence level of $\alpha=0.9$. For all test functions (TP1 to TP9), the dimension of the decision variables (D) was set to 10. To simulate noise, a perturbation radius of 10\% of the domain width was applied to each decision variable with a uniform distribution, utilizing $10^{5}$ noise test samples. All experiments were performed on an Intel(R) Xeon(R) CPU (E5-2678, v3, 2.50GHz) and an RTX A6000 GPU. The RMOEA-UPF algorithm was executed in a Python (3.11.0) environment, while the other comparative algorithms were run on the MATLAB (2023) platform.

Across all algorithms, the total number of real function evaluations was set to 30,000. For all algorithms, the simulated binary crossover (SBX) and polynomial mutation are adopted as the crossover and mutation operators, respectively. Specifically, the distribution index and probability for crossover are set to 20 and 1.0, while the distribution index and probability for mutation are set to 20 and 1/D. Besides, LRMOEA uses its own crossover and mutation operators to generate binary vectors for finding sparse solutions. 

For RMOEA-UPF, the population size was set to 100, the archive capacity was set to 100, the confidence level $\alpha$ was set to 0.9, the elite offspring size was set to 80 for TP1-TP5 and 20 for TP6-TP9. For LRMOEA, the population size was also set to 100. The robustness threshold and the optimization threshold were set to 0.2 and 1.15, respectively, with the remaining settings taken from\cite{10612838}. For MOEA-RE, the population size was set to 100. The maximum perturbation degree was set to 0.1, the coefficient for the threshold was set to 1.5, the sampling size was set to 50, and the remaining settings were refer to \cite{he2019evolutionary}. For RMOEA-SuR, the population size was set to 100. The algorithm was configured with an archive capacity of 7 solutions selected per generation, a precise sampling coefficient of 0.4, and 2 precise samples. The surviving rate threshold was set to 0.9, with the remaining settings referencing \cite{Jiang2025}. For NSGA2-DT1, the population size was set to 100. It was run for 100 iterations, with a neighborhood extent of 0.1 and 2 neighboring points used for mean effective objectives; remaining settings were from \cite{6792442}. For NSGA2, the population size was set to 100, with other configurations based on \cite{996017}.
\subsection{Computational Complexity Analysis}
\begin{table*}[b]
	\caption{The modified Generational Distance (mGD) of each algorithm, calculated as the average shortest distance from its USP set to the $\mathrm{UPF_{global}}$ in the normalized objective space. }
	\begin{center}
		\renewcommand{\arraystretch}{1.5}
		\begin{tabular}{C{0.08}|C{0.135}| C{0.135}| C{0.135} |C{0.135} |C{0.135}| C{0.135}}
			\hline
			Problem&RMOEA-UPF&LRMOEA&MOEA-RE&RMOEA-SuR&NSGA2-DT1&NSGA2
			\\
			\hline
			TP1 &\textcolor{blue}{\makecell{1.246e-2 (±2.0e-4)}} & \textcolor{green}{\makecell{1.211e-2 (±2.5e-4)\\($\approx$)}} & \makecell{1.550e-2 (±2.5e-4)\\($+$)} & \textcolor{red}{\makecell{1.086e-2 (±2.2e-4)\\$-$}} & \makecell{1.339e-2 (±2.8e-4)\\($\approx$)} & \makecell{1.440e-2 (±2.5e-4)\\($\approx$)}\\ \hline
			
			TP2 &\textcolor{green}{\makecell{1.969e-2 (±3.0e-4)}} & \makecell{2.322e-2 (±6.9e-4)\\($+$)} & \makecell{2.724e-2 (±3.0e-4\\($+$)} & \textcolor{red}{\makecell{1.864e-2 (±3.7e-4)\\($\approx$)}} & \textcolor{blue}{\makecell{2.197e-2 (±3.8e-4)\\($+$)}}  &\makecell{2.644e-2 (±2.8e-4)\\($+$)}\\\hline
			
			TP3 & \textcolor{red}{\makecell{1.707e-2 (±1.7e-4)}} & \makecell{2.869e-2 (±3.1e-4)\\($+$)} & \makecell{3.998e-2 (±4.2e-4)\\($+$)} & \textcolor{blue}{\makecell{2.152e-2 (±2.7e-4)\\($+$)}} & \makecell{2.298e-2 (±2.7e-4)\\($+$)} & \textcolor{green}{\makecell{2.143e-2 (±2.7e-4)\\($+$)}}\\\hline
			
			TP4 & \textcolor{red}{\makecell{1.655e-2 (±3.2e-4)}} & \textcolor{green}{\makecell{1.748e-2 (±1.3e-4)\\($\approx$)}} & \makecell{2.108e-2 (±3.6e-4)\\($+$)} & \makecell{1.818e-2 (±3.4e-4)\\($+$)} & \textcolor{blue}{\makecell{1.752e-2 (±3.5e-4)\\($\approx$)}} &\makecell{1.960e-2 (±2.0e-4)\\($+$)}\\\hline
			
			TP5 & \textcolor{red}{\makecell{1.434e-2 (±4.7e-4)}}&\textcolor{green}{\makecell{1.497e-2 (±7.5e-4)\\($\approx$)}} & \makecell{2.011e-2 (±3.3e-4)\\($+$)} & \textcolor{blue}{\makecell{1.603e-2 (±6.7e-4)\\($+$)}} & \makecell{2.053e-2 (±5.1e-4)\\($+$)} & \makecell{1.645e-2 (±5.1e-4)\\($+$)}\\\hline		
			
			TP6 &  \textcolor{red}{\makecell{9.714e-2 (±3.7e-2)}}& \makecell{3.729e-1 (±2.8e-2)\\($+$)} & \makecell{2.867e-1 (±1.5e-2)\\($+$)} &\textcolor{blue}{\makecell{1.387e-1 (±3.9e-2)\\($\approx$)}} & \textcolor{green}{\makecell{1.355e-1 (±2.0e-2)\\($+$)}} &\makecell{1.728e-1 (±2.8e-2)\\$(+)$}\\\hline
			
			TP7 & \textcolor{red}{\makecell{3.752e-2 (±1.1e-3)}}& \makecell{5.085e-1 (±1.4e-3)\\($+$)} & \makecell{1.546e-1 (±1.4e-3)\\($+$)} & \textcolor{blue}{\makecell{6.404e-2 (±9.5e-3)\\($+$)}} & \textcolor{green}{\makecell{6.521e-2 (±7.2e-3)\\($+$)}}& \makecell{6.978e-2 (±1.2e-3)\\($+$)}\\\hline
			
			TP8 &\textcolor{green}{\makecell{1.688e-1 (±2.3e-4)}} & \makecell{1.825e-1 (±7.3e-4)\\($\approx$)} & \makecell{2.611e-1 (±1.2e-3)\\($+$)} & \makecell{3.171e-1 (±5.4e-4)\\($+$)} & \textcolor{red}{\makecell{1.197e-1 (±9.2e-4)\\($-$)}} & \textcolor{blue}{\makecell{3.366e-1 (±3.4e-4)\\($+$)}}\\\hline
			
			TP9 &\textcolor{red}{\makecell{1.149e-1 (±6.8e-3)}} & \makecell{2.220e-1 (±1.4e-2)\\($+$)} & \makecell{3.574e-1 (±1.7e-2)\\($+$)} & \makecell{1.818e-1 (±7.8e-3)\\($+$)} & \textcolor{green}{\makecell{1.331e-1 (±7.3e-3)\\($\approx$)}} & \textcolor{blue}{\makecell{1.656e-1(±1.0e-2)\\($+$)}}\\
			
			\hline
			\multicolumn{2}{c|}{$+/\approx/-$} & \makecell{5/4/0} & \makecell{9/0/0} & \makecell{6/2/1} & \makecell{5/3/1} & \makecell{8/1/0}\\
			\hline
		\end{tabular}
		\begin{tablenotes}   
			\footnotesize             
			\item 
			We report the mean and standard deviation of mGD values over 20 independent runs. The three top-performing algorithms are highlighted using \textcolor{red}{best}, \textcolor{green}{second}, \textcolor{blue}{third}, respectively. Additionally, we use the Wilcoxon Signed Rank Test (with a significance level of 0.05) to verify whether there are significant differences between the comparative algorithms (LRMOEA, MOEA-RE, RMOEA-SuR, NSGA2-DT1, and NSGA2) and RMOEA-UPF, where ``+" indicates significantly better, ``$\approx$" indicates no significant difference, and``-" indicates significantly worse.
		\end{tablenotes}      
		\label{mgd-tp1-9} 
		\vspace{-10pt}  
	\end{center}
\end{table*}
For RMOEA-UPF, let $M$ denote the number of objectives,  the archive capacity be $N_{arc}$ (the core scale parameter), the population size be $N$, and the elite offspring size be $N_{e}$. In general, $N_{arc}$ and $N_{e}$ are of the same order as $N$, so we simplify the analysis by denoting $N=N_{arc}$. The core process driving complexity of RMOEA-UPF is archive updating and final solution selection steps, where each solution in the candidate pool (size proportional to $N$) generates a certain number of USP. Since the total number of solutions in the candidate pool scales with $N$, the total number of USP across all solutions ($T$) is positively correlated with $N$. Generally, $T$ is of the same order as $N$ (i.e., $T\propto N$). Non-dominated sorting of these USP is a critical operation, with a complexity of $O(MT^{2})$. Since $T\propto N$, this simplifies to $O(MN^{2})$. Other steps in the algorithm, such as parent selection, offspring generation, and crowding distance calculation, have lower complexities (e.g., $O(NlogN)$) and do not affect the asymptotic complexity. Overall, the complexity of RMOEA-UPF is $O(MN^{2})$.

For RMOEA-SuR, the complexity of potential solution selection and archive updating processes are \(O(N^2)\) and \(O(N)\) respectively. During the precise sampling step, each relatively optimal solution requires two additional noise disturbance evaluations, resulting in a complexity of \(O(2 \times (N/2)) = O(N)\). The dominant complexity is non-dominated sorting method, with a computational complexity of \(O(MN^2)\), thus making the overall computational complexity of RMOEA-SuR \(O(MN^2)\) \cite{996017}.
Similarly, in MOEA-RE, the dominant complexity is non-dominated sorting method \(O(MN^2)\), the final robust solution selection is performed only once and is not involved in the loop process (complexity \(O(1)\)). Therefore, the computational complexity of MOEA-RE is also \(O(MN^2)\) \cite{8419222}.  For NSGA2, the computational complexity is \(O(MN^2)\) as stated in \cite{996017}. For NSGA2-DT1, let's assume the number of neighboring points within the $\delta$-neighborhood of each solution is H. Thus, the additional complexity for computing average objective functions across the population is  $O(N\cdot H\cdot M)$. Since $H$ is typically a value much smaller than $N$ (e.g., H=3), the additional complexity for robustness evaluation simplifies to$O(N\cdot M)$, which is a lower-order term compared to the core non-dominated sorting complexity. Therefore, the computational complexity of NSGA2-DT1 is also  \(O(MN^2)\) \cite{6792442}.

The computational complexity of LRMOEA is determined by its core operations, including population evolution, archive update, guiding vector generation, and final evaluation \cite{10612838}. LRMOEA adopts selection strategies similar to NSGA-II for population evolution, with a complexity of  \(O(MN^2)\). During archive update step, each solution in the archive is calculated convergence degree (sum of objective values) and deviation (normalized objective variation) involve $O(M)$ computations per solution. With the archive capacity being proportional to $N$, this step has a complexity of $O(N \cdot M)$. In guiding vector generation step, the guiding vector is generated by averaging binary variables in the archive, making this step $O(N)$. In final evaluation, robust solutions are selected using the convergence indicator and weight vector association,  dominated by \(O(N^2)\), which is lower than  \(O(MN^2)\). Thus, the computational complexity of LRMOEA is \(O(MN^2)\).

Therefore, the computational complexity of RMOEA-UPF is \(O(MN^2)\),which is on par with the other state-of-the-art algorithms compared in this study and introduces no additional asymptotic computational burden.

\subsection{Experimental Results}
\begin{figure*}[!h]
	\centering
	
	\subfigure[UPF comparison on TP7]{\includegraphics[width=0.47\linewidth,height=0.27\textheight]{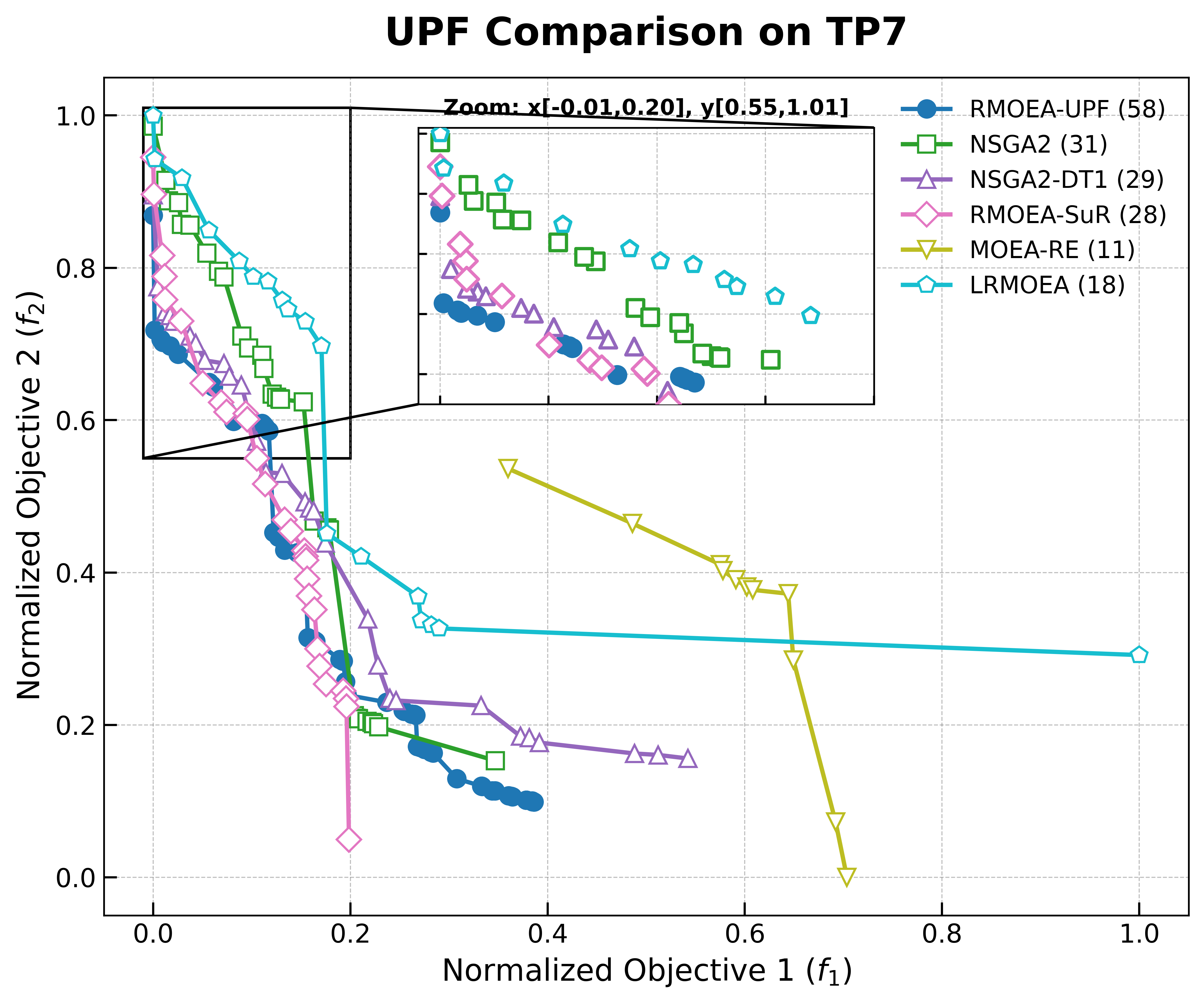}}
	\subfigure[UPF comparison on TP9]{\includegraphics[width=0.47\linewidth,height=0.27\textheight]{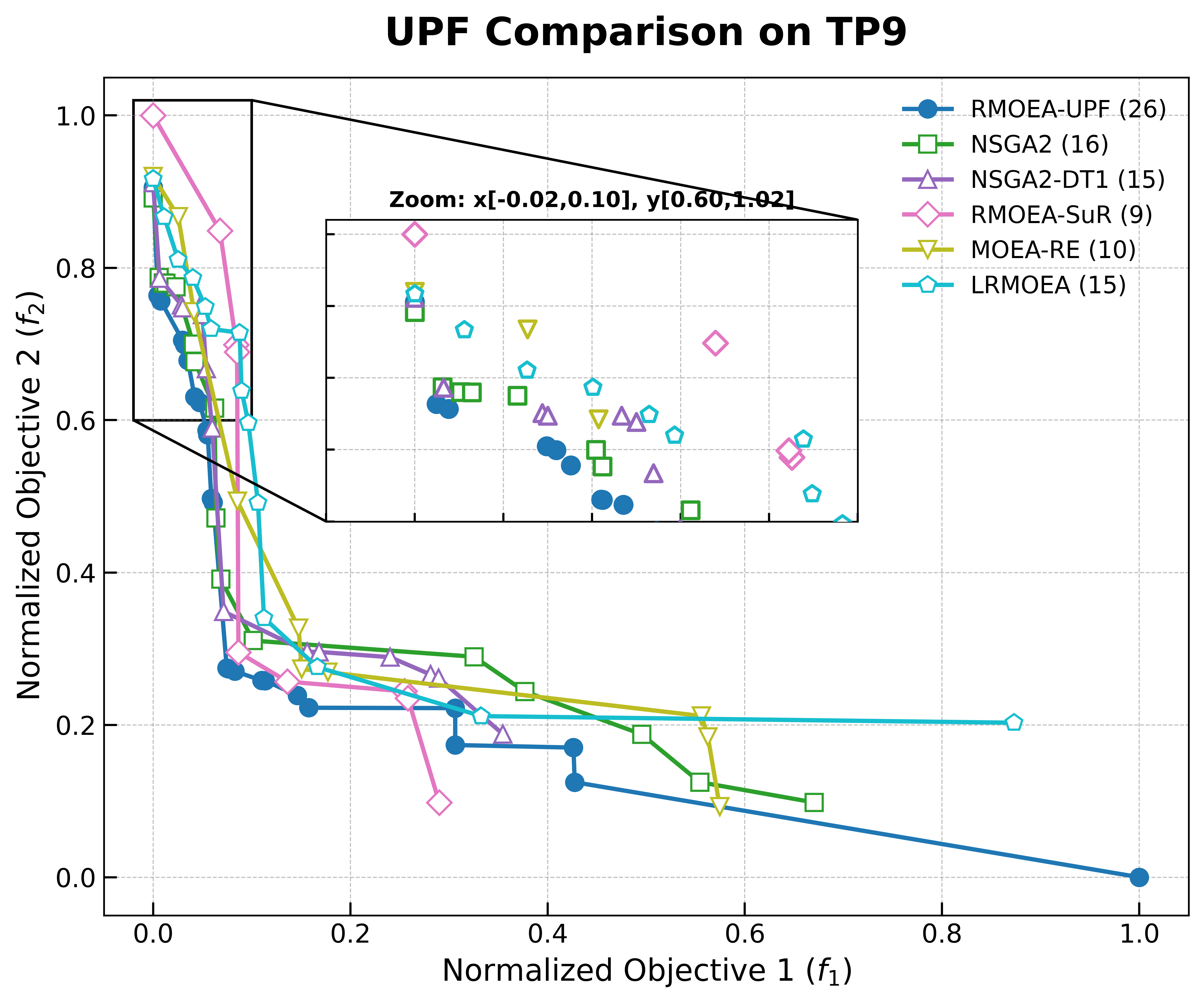}}
	\centering
	\caption{UPF comparison on TP7 and TP9}
	\label{upf-tp7-tp9}
	\vspace{-10pt}
\end{figure*}
The modified Generational Distance (mGD) metric quantifies the average distance between the Uncertainty-related Support Point (USP) set discovered by an algorithm and the global optimal Uncertainty-related Pareto Front ($\mathrm{UPF_{global}}$). A smaller mGD value indicates superior comprehensive performance of the algorithm in terms of robustness and convergence. The results presented in Table \ref{mgd-tp1-9} clearly demonstrate the exceptional performance of RMOEA-UPF across most problems. Among the nine test problems, RMOEA-UPF achieved the optimal mean mGD values in seven cases, specifically TP3, TP4, TP5, TP6, TP7 and TP9. Additionally, the algorithm exhibited strong competitiveness in the remaining two problems (TP2 and TP8), securing the second position. For TP1, its mean mGD value also ranked among the top three.  

Among the comparative algorithms, RMOEA-SuR achieved the best performance on TP1 and TP2. LRMOEA performed prominently on TP1, TP4, and TP5, obtaining the second position in each, but showed relatively weaker performance on other problems. NSGA2-DT1 achieved the optimal result on TP8 and demonstrated favorable performance across other test problems. MOEA-RE and NSGA2 generally yielded higher mGD values than RMOEA-UPF in most cases. Overall, compared with the comparative algorithms, RMOEA-UPF exhibited lower mean mGD values across most problems, with significant advantages particularly on TP3, TP4, TP5, TP6, TP7, and TP9. This indicates that RMOEA-UPF can more effectively search for solutions with superior comprehensive robustness and convergence when addressing multi-objective optimization problems with uncertainty. 

Fig. \ref{upf-tp7-tp9}(a) and Fig. \ref{upf-tp7-tp9}(b) shows the UPF of the final optimal solutions of various algorithms on the TP7 and TP9 test functions in one of the experiments (the numbers enclosed in parentheses to the right of the legend items indicate the number of the points on each algorithm's UPF), respectively.
For more comprehensive visualization information, please refer to the \textbf{Appendix \ref{appendix-a}}. We have visualized the mappings of the final solution sets obtained by all algorithms on the TP1-TP9 test functions in the objective space, as well as the UPF after applying noise perturbations following a uniform distribution. 

We also conducted comparative experiments on the IGD metric. Due to space constraints in the paper, please refer to \textbf{Appendix \ref{appendix-b}} for details. Combining the results from the mGD and IGD metrics, we can conclude that the RMOEA-UPF algorithm exhibits significant advantages in comprehensive performance (convergence and robustness) when addressing multi-objective optimization problems with uncertainty.

\subsection{Real-world Problem: Optimal Control of Greenhouse Microclimate}
To validate the effectiveness and practical applicability of the proposed RMOEA-UPF algorithm, we apply it to a real-world robust multi-objective optimization problem: the optimal control of a greenhouse microclimate as shown in Fig. \ref{greenhouse} In Section \ref{introduction}).

The optimal setpoints for regulating greenhouse microclimate variables $X$ (temperature, $CO_2$ concentration, and light intensity) are derived from a multi-objective optimization aimed at maximizing crop yield ($f_1$) and minimizing regulation costs ($f_2$), such as expenses related to  $CO_2$ supplementation, artificial lighting, and temperature control energy consumption. Since crop yield is proportional to the net photosynthetic rate, the yield maximization objective is transformed into maximizing the net photosynthetic rate of the crop. Therefore, the multi-objective optimization function can be defined as follows (since a higher photosynthetic rate is desirable, a negative sign is added before the term):

\begin{equation}  
	\begin{split}   
		\min\quad &f_{1}(x) = -P(C, I,T) \\   
		&f_{2}(x) = Q(C,I,T) = \left( \alpha_1 Q_{C} + \alpha_2 Q_I +\alpha_2 Q_T\right) \\   
		\text{s.t.}\quad &
		\begin{cases}     
			C_{\min} \le C \le C_{\max} \\    
			I_{\min} \le I \le I_{\max}  \\
			T_{\min} \le T \le T_{\max}     
		\end{cases}  
	\end{split} 
\end{equation}
where \( P \) refers to the photosynthetic rate, \( Q \) denotes the economic cost of regulation, \( C \) represents the $CO_2$ concentration, \( I \) stands for light intensity, and \( T \) indicates temperature. 

Given that temperature is assumed to be maintained within an optimal range by a dedicated climate control system—thus treated as a fixed parameter in the photosynthetic rate model with distinct coefficient factors—we adopt the rectangular hyperbolic correction model for photosynthetic rate proposed in \cite{Ye2010Comparison} as our first optimization objective $f_{1}(x)$  in Eq. (\ref{real-problem}). This model is dependent on the two decision variables of $CO_2$ concentration ($x_1$) and light intensity ($x_2$). We utilize the regulation cost model (primarily derived from $CO_2$ supplementation and artificial lighting) constructed in \cite{wen2022research} as second objective $f_{2}(x)$ in Eq. (\ref{real-problem}). 

\begin{equation}  \label{real-problem}  
	\begin{split}   
		\min\quad &  f_{1}(x) =-\left ( x_{1} \frac{1 - \beta_c x_1}{1 + \gamma_c x_1}\right ) \left ( \frac{1 - \beta_i x_2}{1 + \gamma_i x_2} x_{2}\right  ) \\   
		& f_{2}(x) = \alpha _{1} V x_1 + \alpha _{2}  x_2 \\
	\end{split} 
\end{equation}
where $x_1$ represents $CO_2$ concentration ($\mathrm{\mu mol/mol}$), $x_2$ represents light intensity ($\mathrm{\mu mol/mol}$), \( \beta_c, \gamma_c, \beta_c, \gamma_i \) are photosynthetic rate factors, $\alpha _{1}$ is the cost coefficient for $CO_2$ supplementation, \( V \) denotes the total volume of the greenhouse ($\mathrm{m^3}$) and $\alpha _{2}$ is the coefficient for supplemental lighting. 

However, in practical applications, the greenhouse microclimate is highly susceptible to influences from outdoor weather. Due to the difficulty in accurately predicting outdoor weather, inevitable uncertain deviations $\delta$ arise between the actual and predicting values of the indoor microclimate. These deviations can be regarded a ``noise disturbances''. Therefore, the above optimization problem represents a practical robust multi-objective optimization problem where decision variables are subject to noise disturbances.

To simulate realistic operational fluctuations, we set the perturbation radius for each decision variable to 5\% of its domain width. For the greenhouse problem, the primary decision variables $x_1$ and $x_2$ are constrained to the ranges [0, 1200] and [0, 1500] respectively, while the remaining variables are constrained to [0, 1]. The objective model parameters are set as follows: $\beta_c$=2.12, $\gamma_c$=5.75, $\beta_i=6.34 \times 10^{-5}$, $\gamma_i=3.59 \times 10^{-3}$; \( V \)=1918; $\alpha _{1}$=0.345, and $\alpha _{2}$=$6 \times 10^{-4}$. For RMOEA-UPF, the parameters were set as follows: a population size of 100, an archive capacity of 100, 30 elite offspring selected per generation, and a confidence level $\alpha$ of 0.9. The total number of real function evaluations was limited to 20,000. For consistency and comparability, the parameters of all competing algorithms were configured in accordance with the specifications provided in Section \ref{settings}.
\begin{table}[!h]
	\caption{The mGD and IGD metrics of RMOEA-UPF and other comparative algorithms in normalized objective space.}
	\begin{center}
		\renewcommand{\arraystretch}{1.5}
		\begin{tabular}{l|c|c}
			\hline
			Algorithm & mGD & IGD \\
			\hline
			
			RMOEA-UPF & \textcolor{red}{\makecell{1.315e-2 (±8.3e-4)}} & \textcolor{red}{\makecell{9.914e-3 (±1.2e-3)}} \\
			NSGA2     & \makecell{8.927e-1 (±7.9e-3)} & \makecell{1.752e-1 (±1.4e-2)} \\
			NSGA2-DT1 & \textcolor{green}{\makecell{1.642e-2 (±1.3e-4)}} & \textcolor{blue}{\makecell{2.306e-2 (±2.4e-3)}} \\
			RMOEA-SuR & \textcolor{blue}{\makecell{4.403e-2 (±4.6e-4)}} & \textcolor{green}{\makecell{2.152e-2(±4.5e-3)}} \\
			MOEA-RE   & \makecell{9.830e-2 (±7.3e-4)} & \makecell{1.738e-1 (±1.3e-2)} \\
			LRMOEA    & \makecell{4.500e-2 (±1.4e-4)} & \makecell{1.741e-1 (±1.2e-2)} \\
			\hline
		\end{tabular}
		\begin{tablenotes}   
			\footnotesize             
			\item 
			We report the mean and standard deviation of IGD values over 20 independent runs. The three top-performing algorithms are highlighted using \textcolor{red}{best}, \textcolor{green}{second}, \textcolor{blue}{third}, respectively. 
		\end{tablenotes}         
		\label{Real-world-results}
		\vspace{-10pt}
	\end{center}
\end{table}

	\begin{figure}[!h]
		\centering
		\includegraphics[scale=0.4]{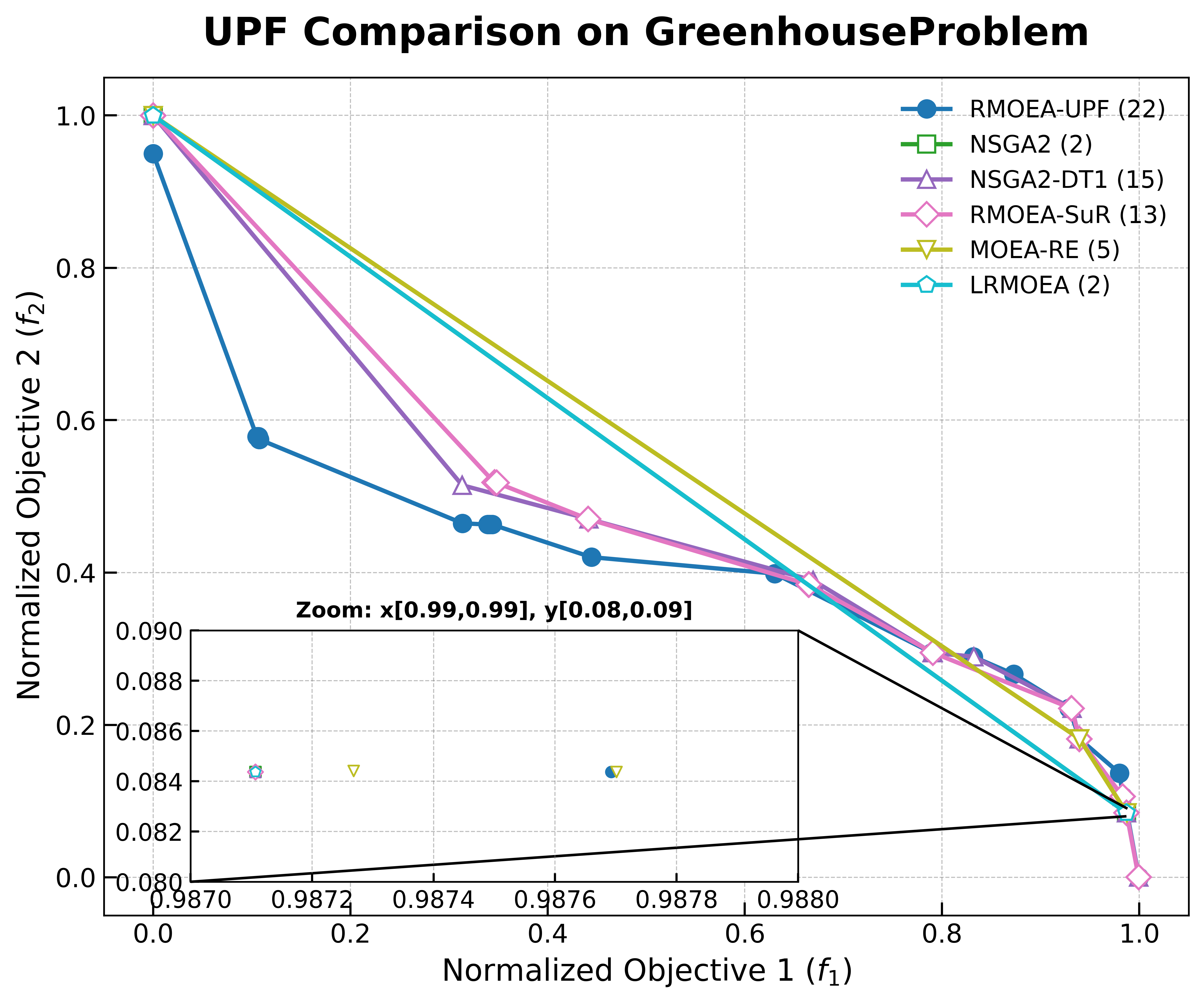}
		\caption{UPF comparison on optimal control of greenhouse microclimate}
		\label{Real-world}
		\vspace{-10pt}
	\end{figure}
Table \ref{Real-world-results} presents the mean and standard deviation of two metrics (including the modified Generational Distance (mGD) and the Inverted Generational Distance (IGD) of the Uncertainty-related Pareto Fronts (UPF) of the solution sets generated by various algorithms on this greenhouse environmental optimal control problem, obtained from 20 independent runs of comparative experiments. From Table \ref{Real-world-results}, the RMOEA-UPF algorithm demonstrates a significant competitive advantage when addressing multi-objective optimization problems under uncertainty. On the mGD metric, RMOEA-UPF achieved the best performance with a mean value of 1.315e-2, outperforming all other algorithms. Similarly, RMOEA-UPF also ranked first on the IGD metric with a mean of 9.914e-3, indicating its superior performance in both robustness and convergence. The NSGA2-DT1 algorithm secured second place on the mGD metric and third on the IGD metric. RMOEA-SuR, on the other hand, ranked third on mGD, and showed second performance on the IGD metric. Other algorithms, such as NSGA2, MOEA-RE, and LRMOEA, exhibited relatively weaker performance, with their mGD and IGD mean values consistently ranking below the top three. Notably, on the IGD metric, their mean values were in the range of 1.741e-1 to 1.752e-1, indicating a less effective approach to handling uncertainty. Fig. \ref{Real-world} shows the UPF of the final optimal solutions obtained by various algorithms on real-world problems in one of the experiments. It can be observed that RMOEA-UPF achieves the best-performing Uncertainty-related Pareto Front (UPF).
	
In summary, RMOEA-UPF showcases outstanding competitiveness across both the mGD and IGD performance metrics. Its dominant position, particularly on the mGD metric, validates its effectiveness in finding highly converged and robust solution sets in real, complex and uncertain environments.

\subsection{Parameter Sensitivity Analysis Experiment}
\begin{figure*}[t]
	\centering
	
	\subfigure[UPF comparison for different $alpha$]{\includegraphics[width=0.47\linewidth,height=0.27\textheight]{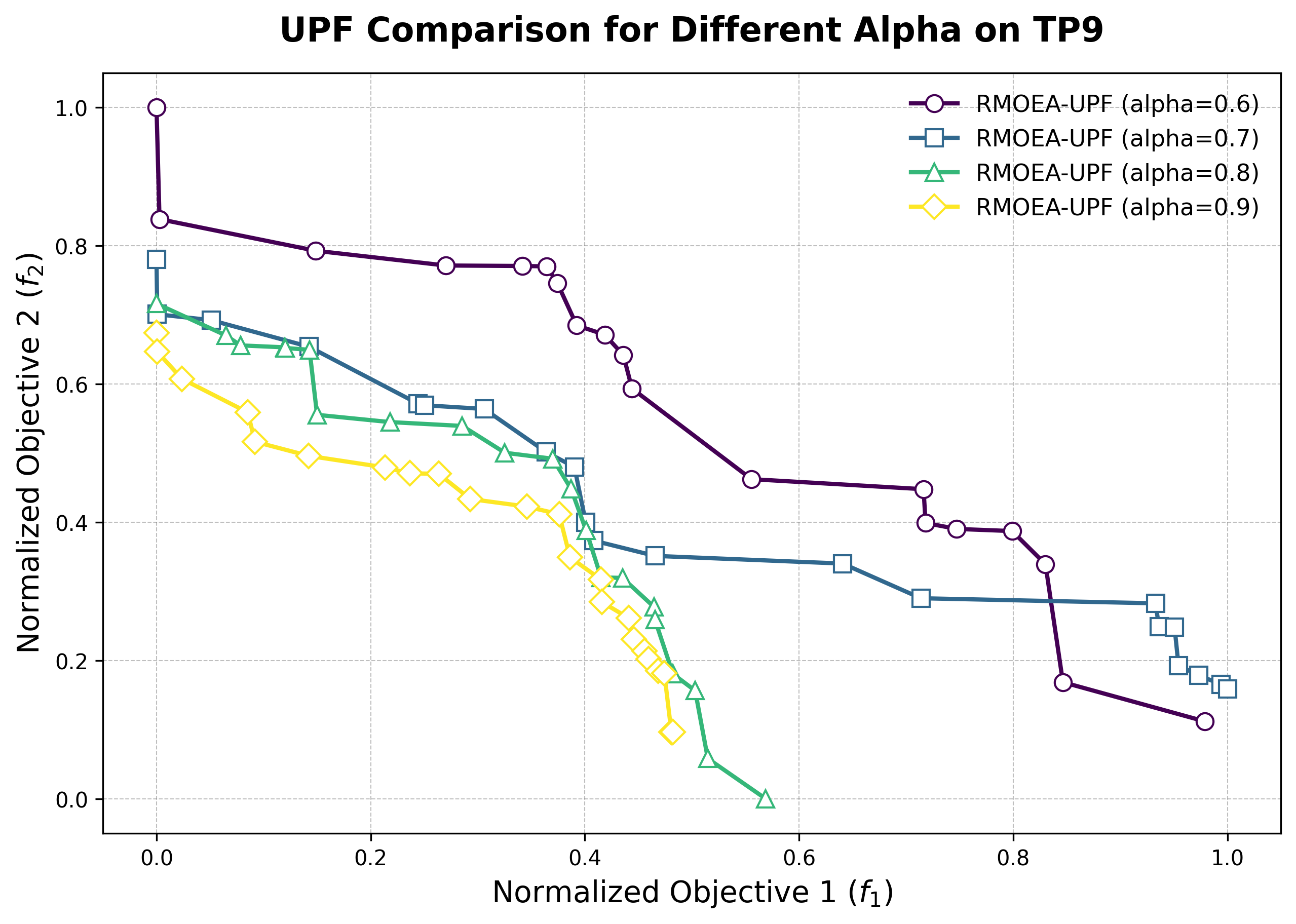}}
	\subfigure[Convergence comparison for different $alpha$]{\includegraphics[width=0.47\linewidth,height=0.27\textheight]{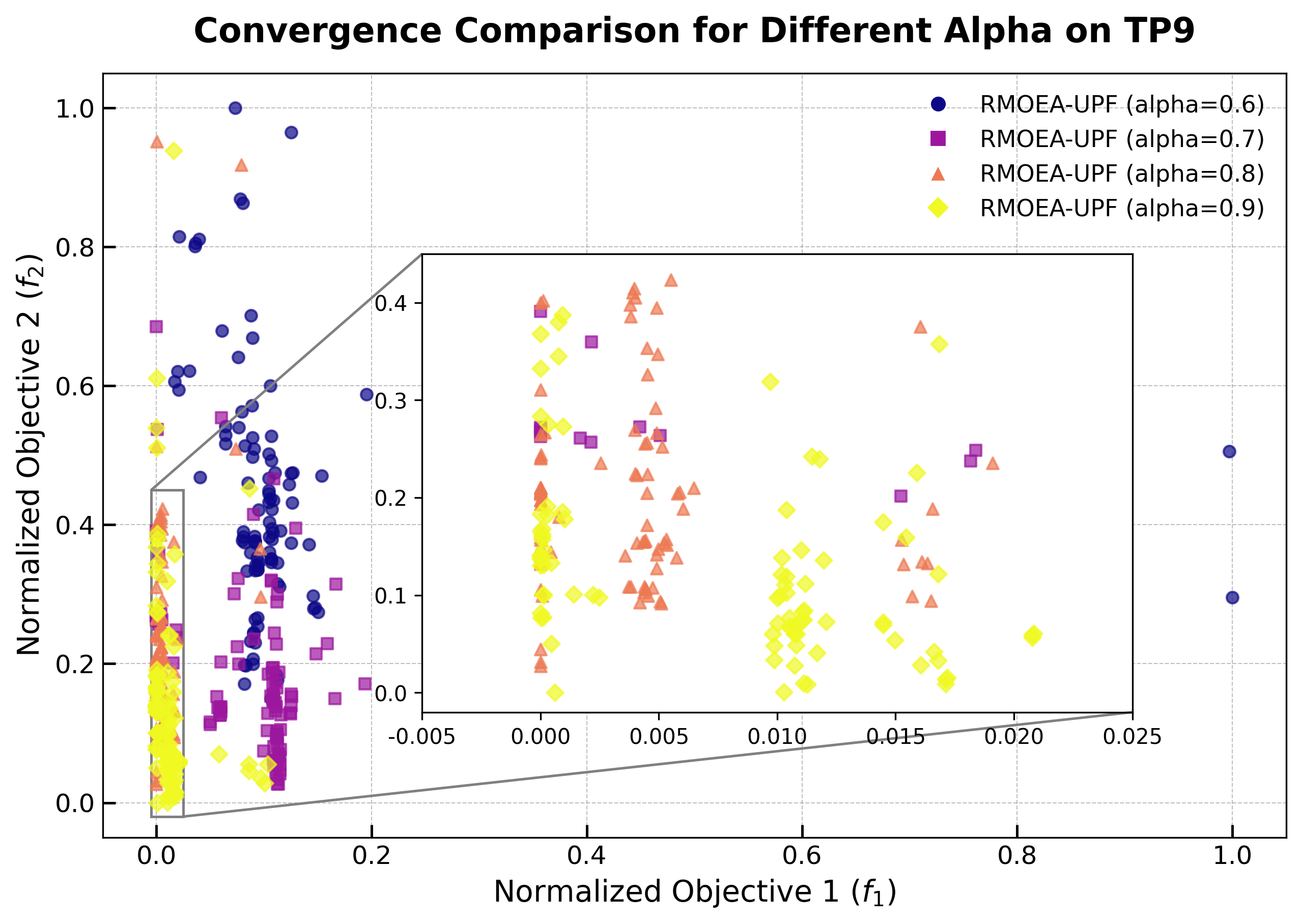}}
	\centering
	\caption{Parameter sensitivity analysis experiment of $alpha$}
	\label{parameter-alpha}
	\vspace{-10pt}
\end{figure*}
The confidence level $\alpha$ is a fundamental parameter in the Uncertainty-related Pareto Front (UPF) framework, as it dictates the probability with which a solution's performance under noise will be no worse than its Uncertain $\alpha$-Support Points (USP). Unlike traditional robust optimization, where robustness is defined by a single metric, $\alpha$ within the UPF framework simultaneously governs the trade-off between robustness and convergence. A higher $\alpha$ compels the algorithm to seek solutions that perform well under nearly all noise perturbations, thereby prioritizing robustness.

To explore the influence of this parameter, we conducted a sensitivity analysis using the TP9 test function. We varied $\alpha$ across values of 0.6, 0.7, 0.8, and 0.9. The algorithm was run with a maximum of 20,000 evaluations, with all other parameters held constant. We first compared the UPF obtained in the normalized objective space after exposing the solution sets to 10,000 random noise instances (Fig. \ref{parameter-alpha}(a)). The results demonstrate that as $\alpha$ increases, the quality of the UPF improves, which aligns with our theoretical definition.

Furthermore, we examined the final solution sets' mapping in the objective space (Fig. \ref{parameter-alpha}(b)). The results for $\alpha = 0.8$ and $\alpha = 0.9$ show a marked shift of the objective points to the left compared to $\alpha = 0.6$ and $\alpha = 0.7$. This indicates a significant improvement in convergence, validating that the $\alpha$ value in the UPF framework directly determines the combined performance of both robustness and convergence (rather than defining robustness in isolation). When $\alpha$ is set to a larger value, both robustness and convergence are enhanced. 

Furthermore, we have also conducted sensitivity analyses with different parameter settings for the archive capacity ($N_{arc}$) and elite offspring size ($N_e$), respectively. For detailed information, please refer to the \textbf{Appendix \ref{appendix-c}}. 
\section{Conclusion}\label{conclusion} 
In this paper, we analyze the limitations of existing research when handling uncertain multi-objective optimization problems and provide a systematic framework to  balance solution convergence and robustness. We innovatively propose the concept of the ``Uncertainty-related Pareto Front" (UPF), which equally considers robustness and convergence in line with the definition of uncertain multi-objective optimization problems. The proposal of the UPF lays a theoretical foundation for the development of high-efficiency search optimization methods for uncertain problems. Through this innovation, we extend the study of robust multi-objective optimization from the pursuit of a singular preferred solution to a more efficient UPF-based population search optimization approach.  Moreover, we develop a robust multi-objective optimization algorithm based on UPF (RMOEA-UPF) under noise perturbation in decision variables. This algorithm employs an innovative archive updating strategy to effectively approximate and identify the UPF during the evolutionary iteration process. By continuously guiding the population's UPF toward the optimal direction, the algorithm achieves an efficient population-based search for robust optimal solutions while equally considering robustness and convergence during the optimization process. 

To validate the effectiveness of the proposed algorithm, we selected nine robust multi-objective optimization benchmark test problems and conducted comparative experiments with typical algorithms developed for such problems. The experimental results demonstrate that RMOEA-UPF exhibits exceptional robustness and convergence properties when solving multi-objective optimization problems under noisy conditions. Notably, the algorithm maintains consistently superior performance across all test scenarios, showcasing its remarkable generalization capability. The superior performance of RMOEA-UPF fundamentally originates from its innovative design framework that simultaneously addresses robustness and convergence through the proposed UPF. The consistent effectiveness demonstrated in our experiments confirms that UPF represents an effective paradigm for handling real-world optimization problems subject to variable noise disturbances.

Despite the promising results, our investigation has been specifically focused on mitigating the impact of uncertainty in decision variables—a core challenge in many real-world optimization problems. While this provides a deep analysis for this context, we recognize that other forms of uncertainty, such as noisy objective function measurements or uncertain model parameters, are vital research directions that could build upon the UPF framework in the future. Furthermore, while RMOEA-UPF has proven effective, its scalability to problems with very high-dimensional decision spaces (large-scale RMOPs) or a large number of objectives (many-objective RMOPs) presents promising avenues for future research. For large-scale problems, future work should focus on developing strategies to maintain search efficiency within the UPF framework. For many-objective problems, it will be crucial to investigate alternative selection mechanisms that can effectively handle a high number of conflicting objectives in conjunction with the UPF's probabilistic guarantees. Additionally, integrating the UPF framework with advanced surrogate models presents a promising research direction, particularly for computationally expensive problems where real function evaluations are limited. Moreover, we note the scarcity of test functions specifically designed to challenge algorithms that, like ours, treat robustness and convergence co-equally. The development of such benchmarks will be a focal point for future research to further advance the field. Ultimately, the UPF framework offers a principled and extensible foundation for developing next-generation algorithms capable of tackling complex optimization challenges in an uncertain world.

\bibliographystyle{IEEEtran}
\bibliography{reference}

\newpage
\onecolumn
\appendices
\section{Comparative Experimental Result Diagrams for TP1-TP9 problems}\label{appendix-a}
For each test problem, all algorithms were independently executed 20 times. To provide a representative visualization, we selected the results from a single, typical run for plotting.  For each test problem, we present two distinct figures to comprehensively illustrate the performance of the compared algorithms. The first figure (on the left), titled ``Objective Space Mapping," displays the final solution sets obtained by each algorithm, evaluated deterministically without noise. In the legend of each plot, the number displayed next to the algorithm's name indicates the total number of solutions in its final obtained set. 

The second figure (on the right), titled ``UPF Comparison," illustrates the robustness of these solution sets. To generate this, we applied a pre-generated dataset of $10^5$ uniform noise perturbations (radius of 10\% of the domain width) to each final solution set and then calculated the corresponding Uncertainty-related Pareto Front (UPF). This plot visualizes the effective performance front of each algorithm under uncertainty. In the legend of these UPF plots, the number next to the algorithm's name denotes the number of USP points that constitute its final non-dominated front.

\begin{figure*}[!h]
	\centering
	
	\subfigure[Mapping of the solution set in the objective space]{\includegraphics[width=0.47\linewidth,height=0.27\textheight]{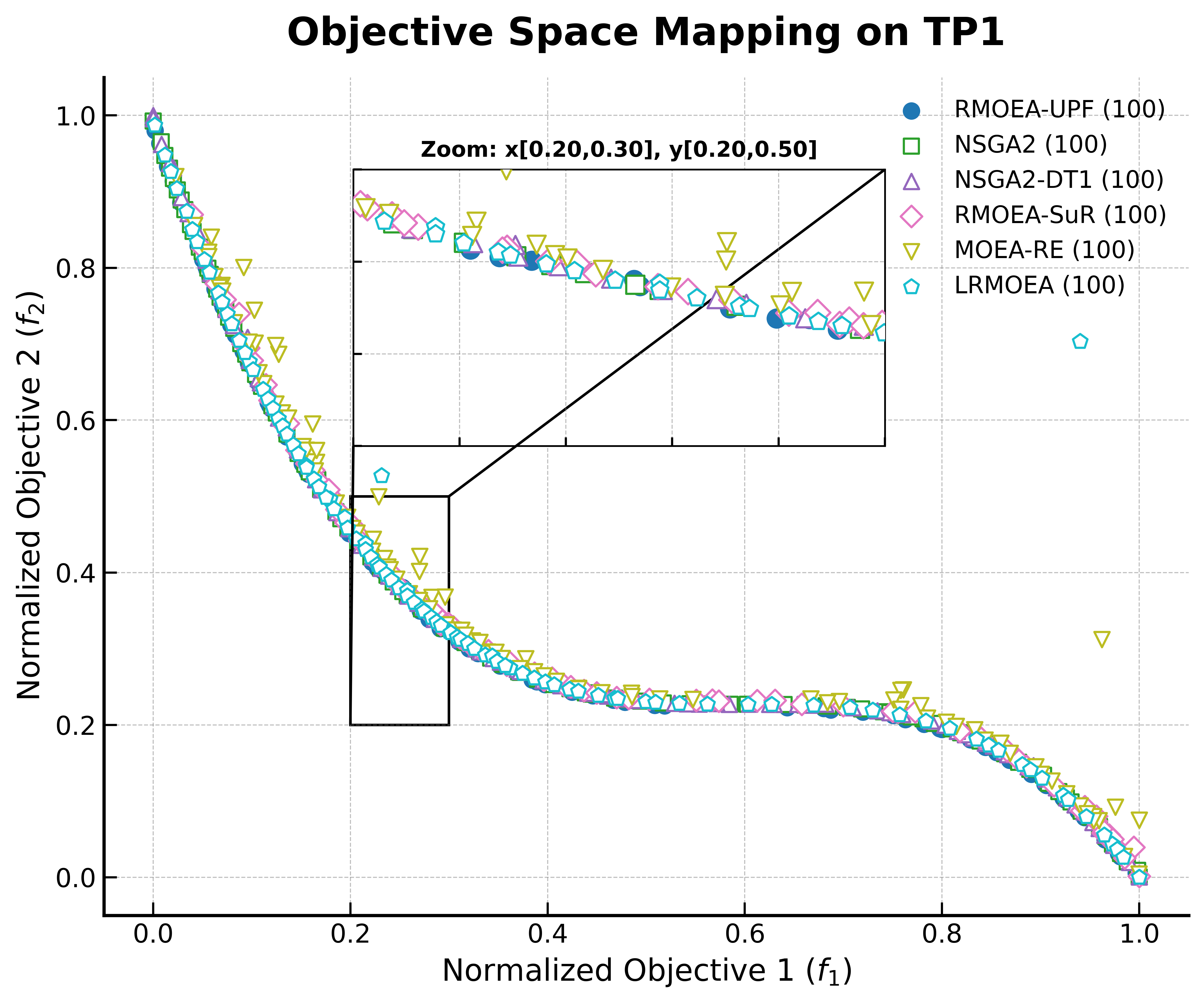}}
	\subfigure[Uncertainty-related Pareto Front]{\includegraphics[width=0.47\linewidth,height=0.27\textheight]{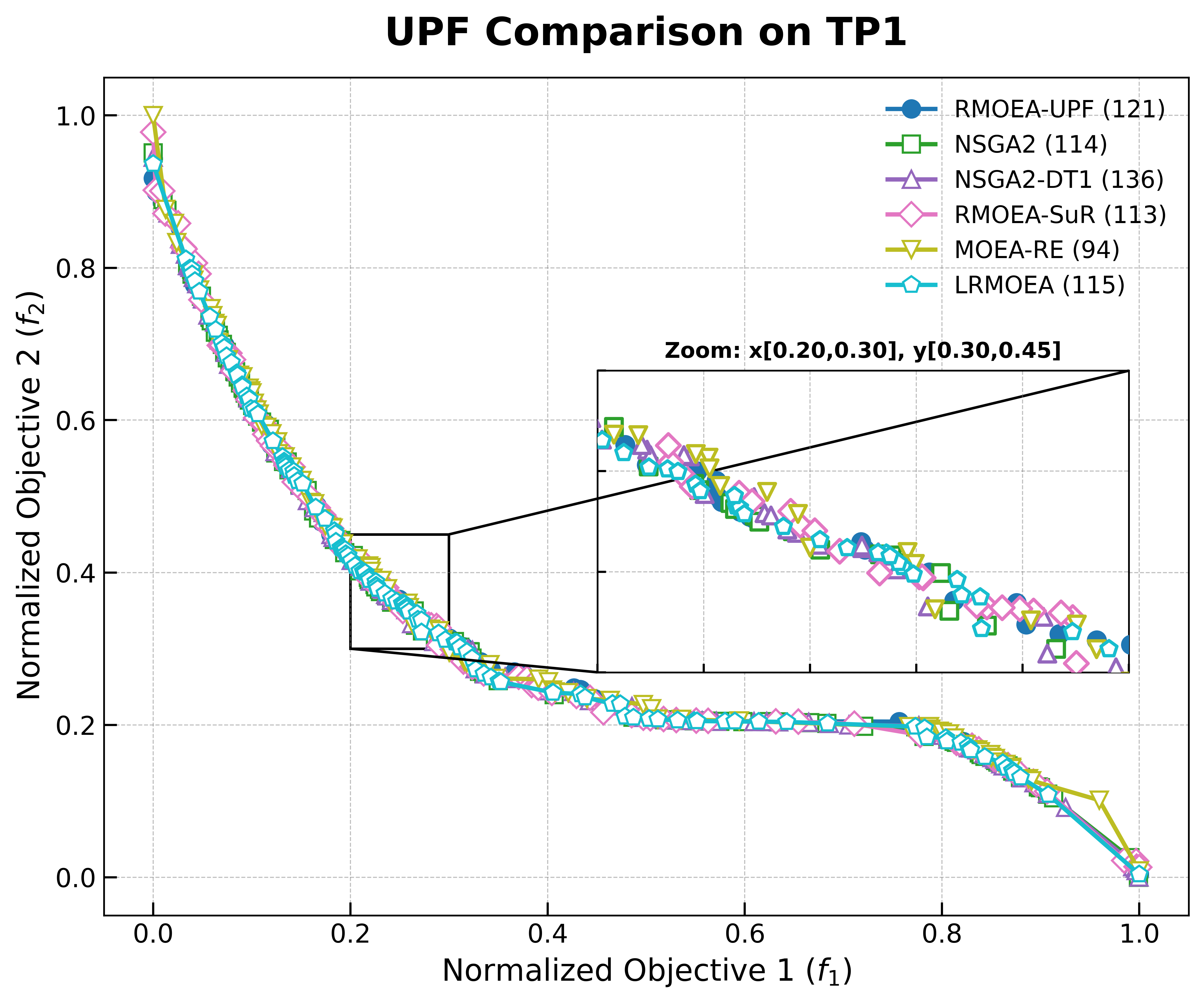}}
	\centering
	\caption{TP1 problem}
	\label{TP1}
	\vspace{-20pt}
\end{figure*}
\begin{figure*}[!h]
	\centering
	
	\subfigure[Mapping of the solution set in the objective space]{\includegraphics[width=0.47\linewidth,height=0.27\textheight]{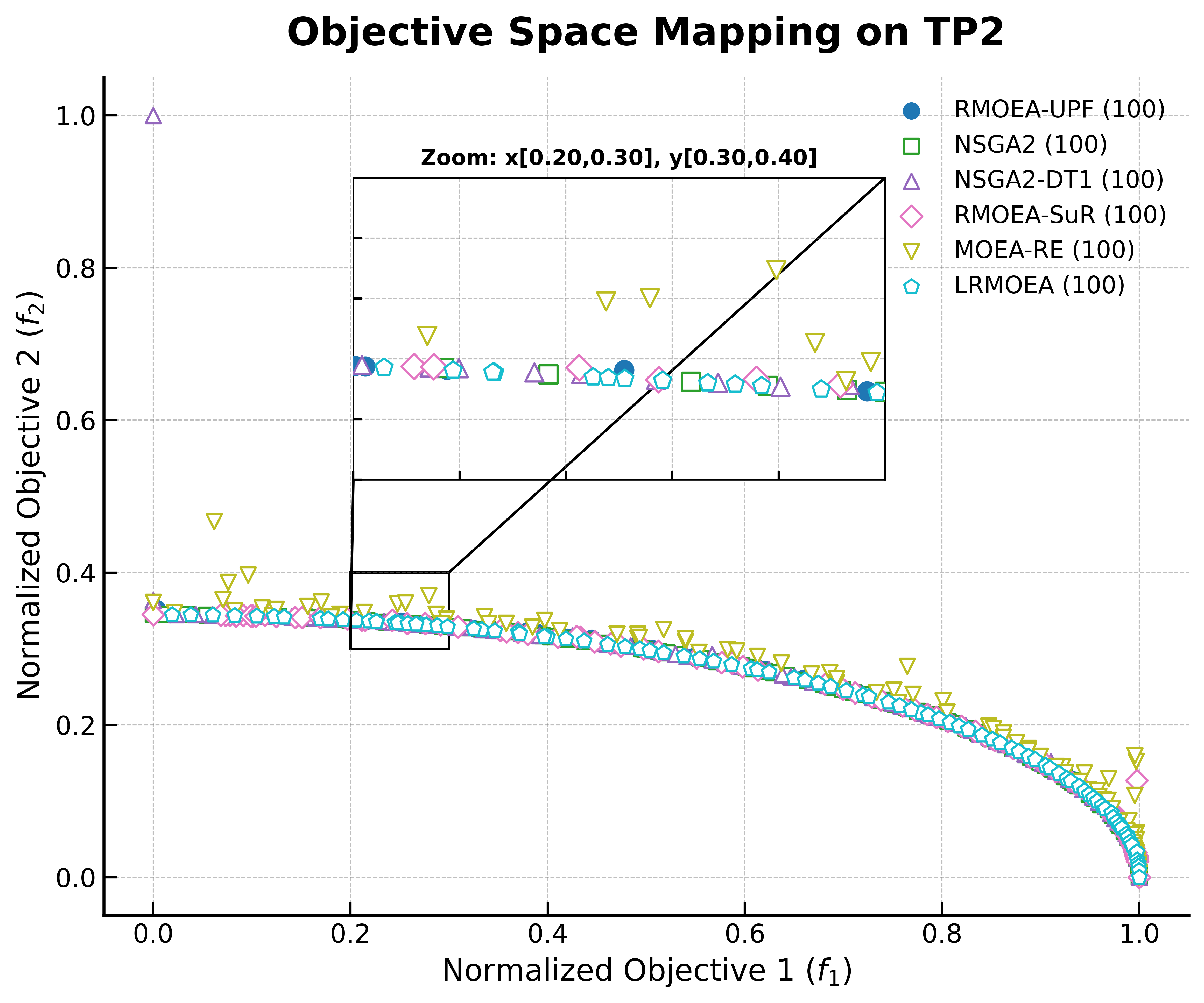}}
	\subfigure[Uncertainty-related Pareto Front]{\includegraphics[width=0.47\linewidth,height=0.27\textheight]{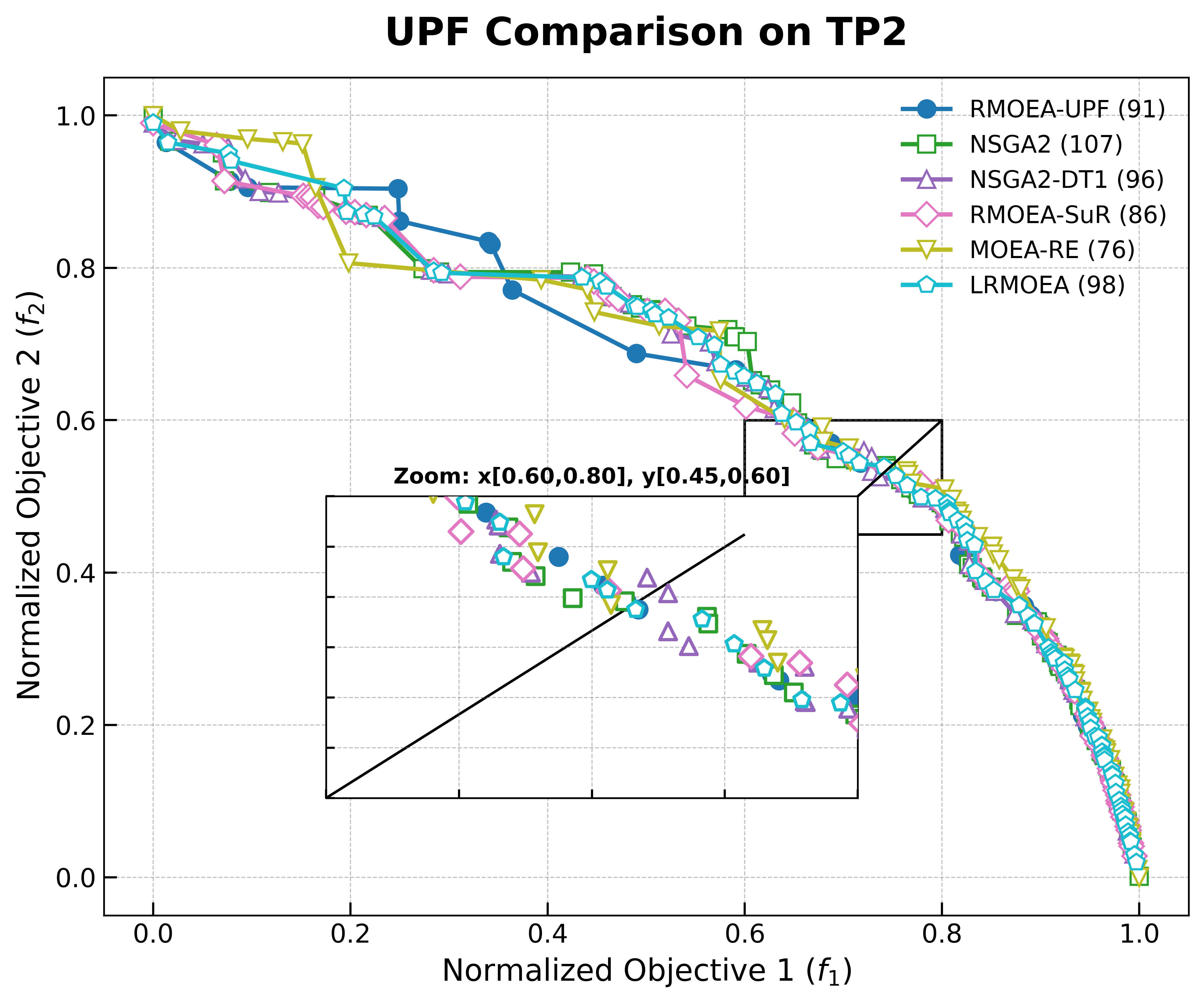}}
	\centering
	\caption{TP2 problem}
	\label{TP2}
	\vspace{-20pt}
\end{figure*}

\begin{figure*}[!h]
	\centering
	
	\subfigure[Mapping of the solution set in the objective space]{\includegraphics[width=0.47\linewidth,height=0.27\textheight]{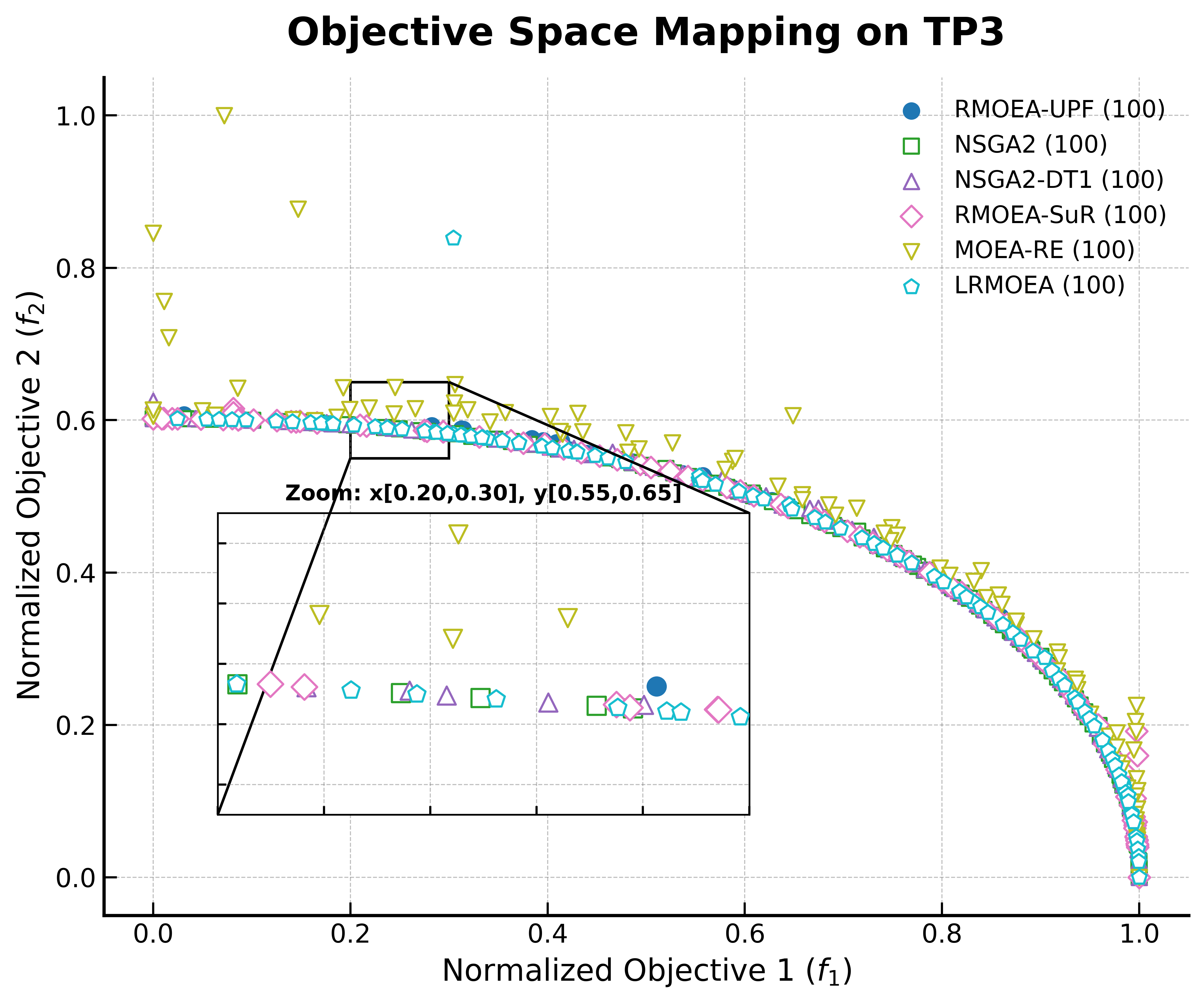}}
	\subfigure[Uncertainty-related Pareto Front]{\includegraphics[width=0.47\linewidth,height=0.27\textheight]{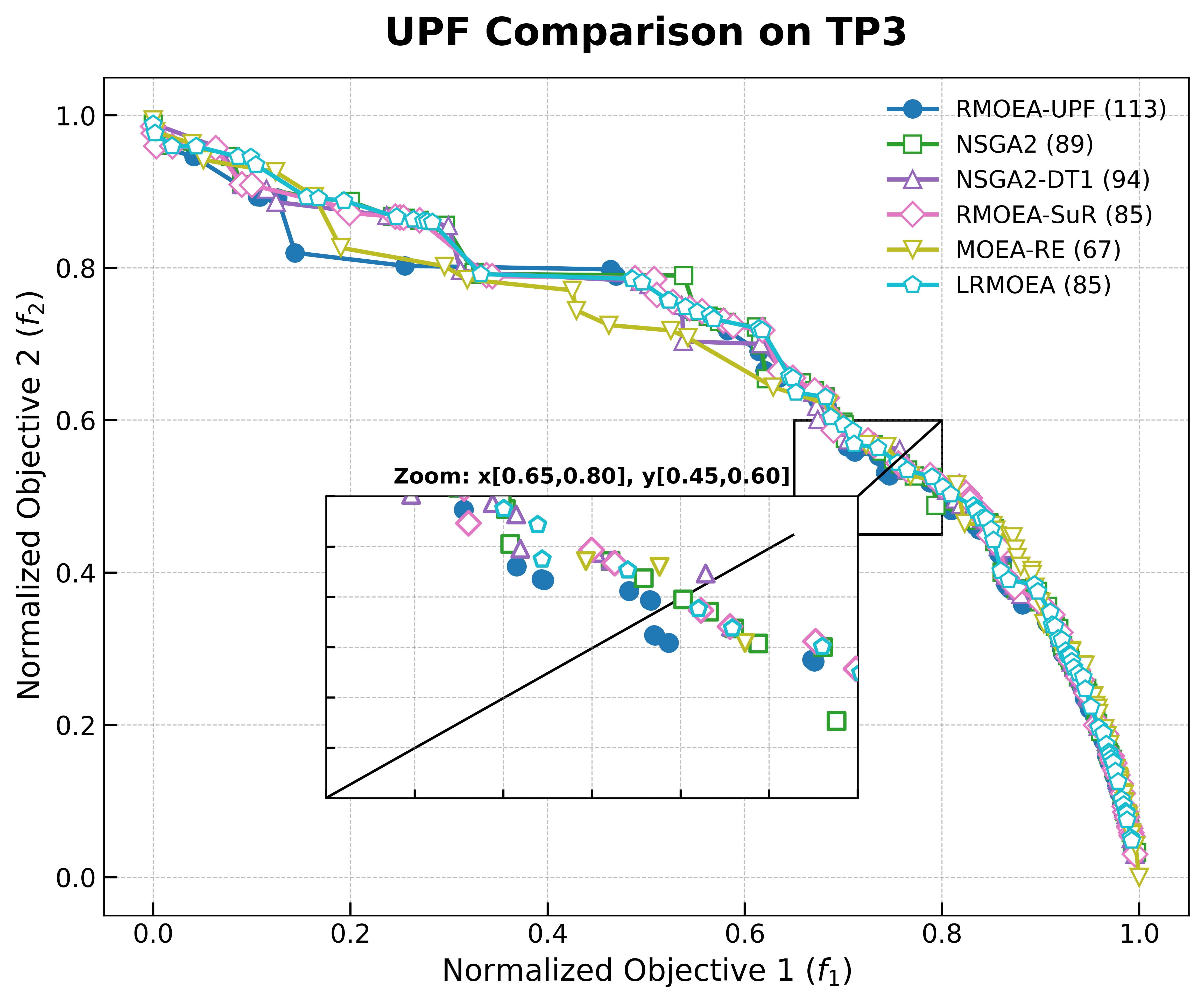}}
	\centering
	\caption{TP3 problem}
	\label{TP3}
	\vspace{-20pt}
\end{figure*}

\begin{figure*}[!h]
	\centering
	
	\subfigure[Mapping of the solution set in the objective space]{\includegraphics[width=0.47\linewidth,height=0.27\textheight]{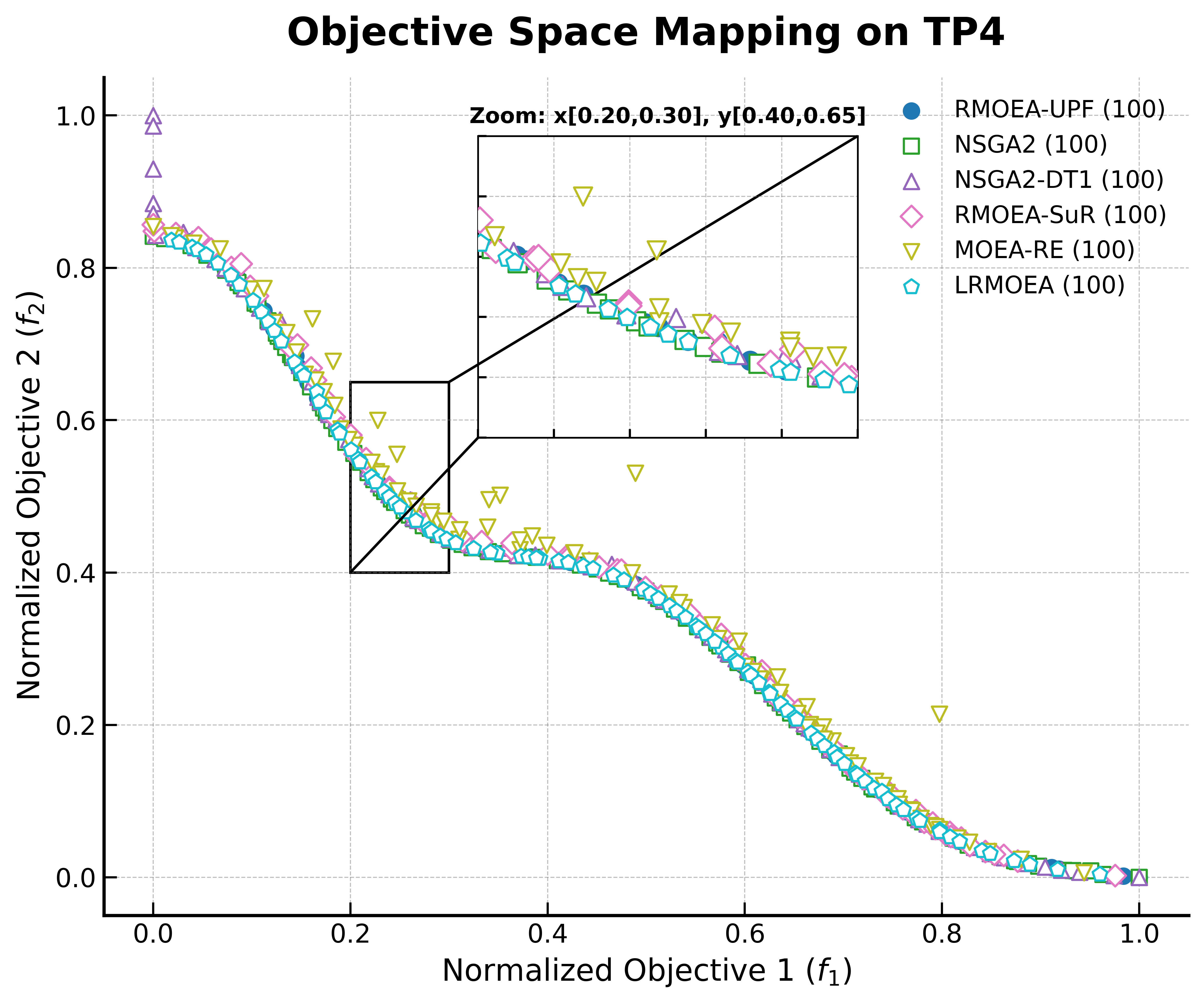}}
	\subfigure[Uncertainty-related Pareto Front]{\includegraphics[width=0.47\linewidth,height=0.27\textheight]{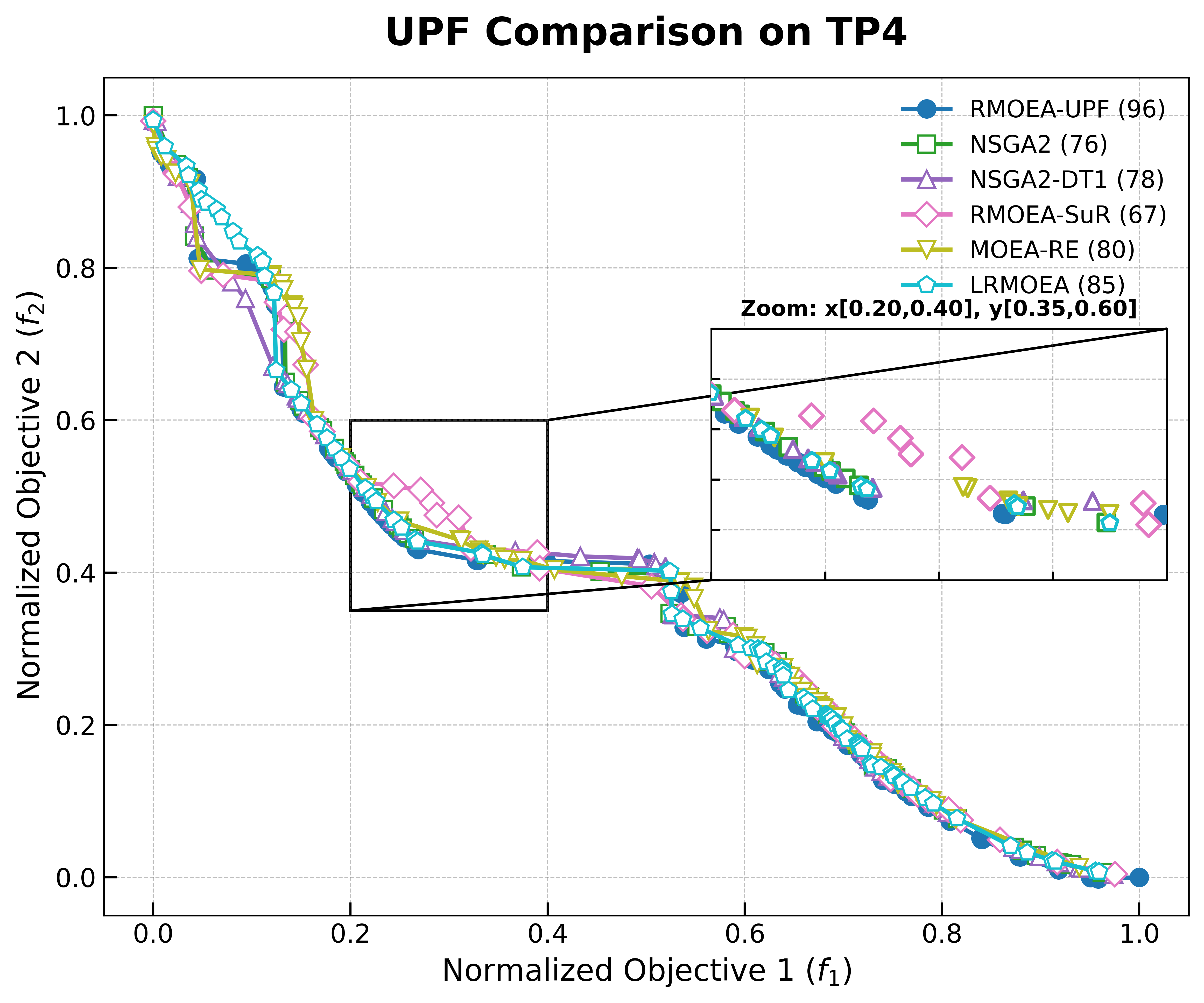}}
	\centering
	\caption{TP4 problem}
	\label{TP4}
	\vspace{-20pt}
\end{figure*}
\begin{figure*}[!h]
	\centering
	
	\subfigure[Mapping of the solution set in the objective space]{\includegraphics[width=0.47\linewidth,height=0.27\textheight]{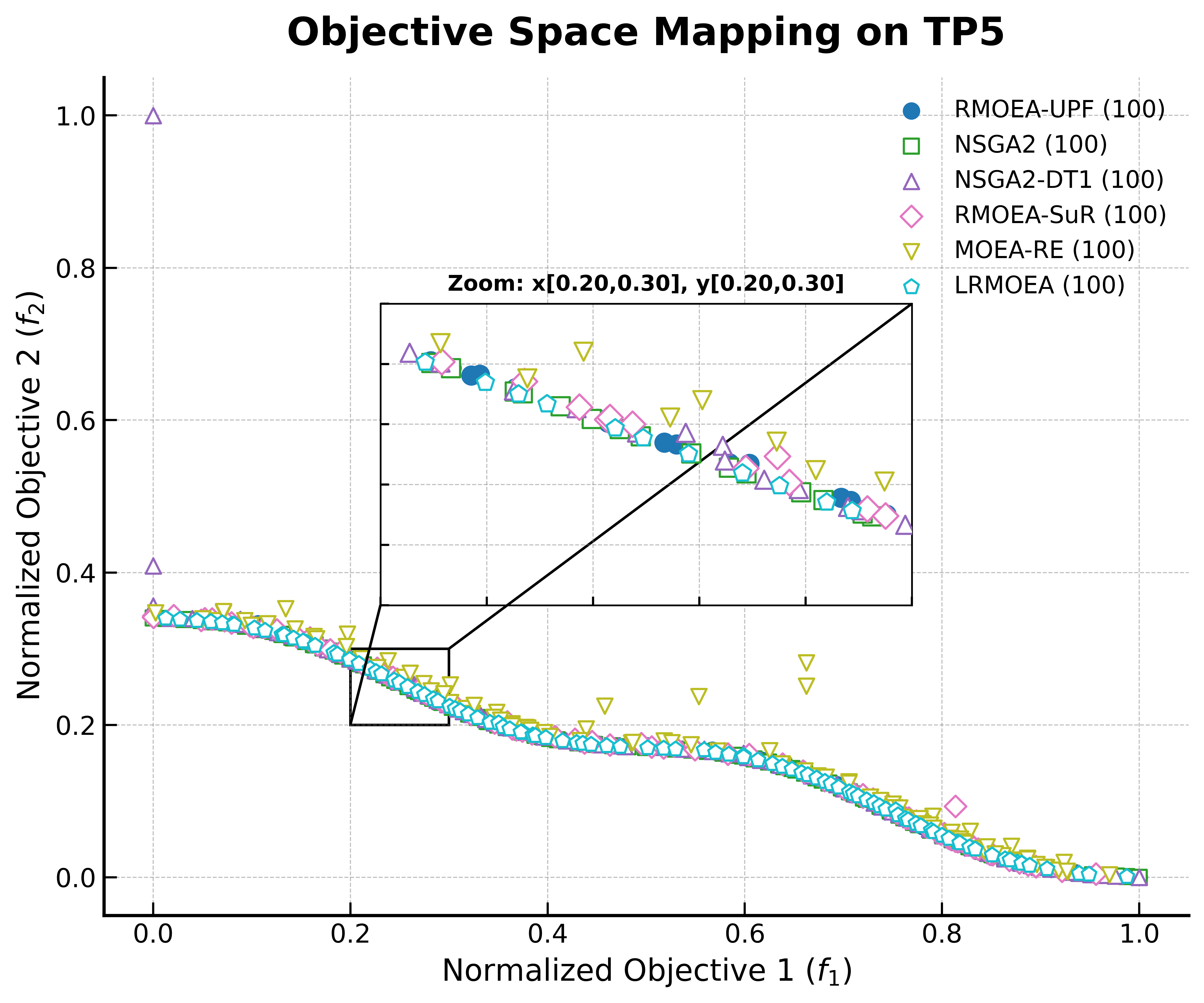}}
	\subfigure[Uncertainty-related Pareto Front]{\includegraphics[width=0.47\linewidth,height=0.27\textheight]{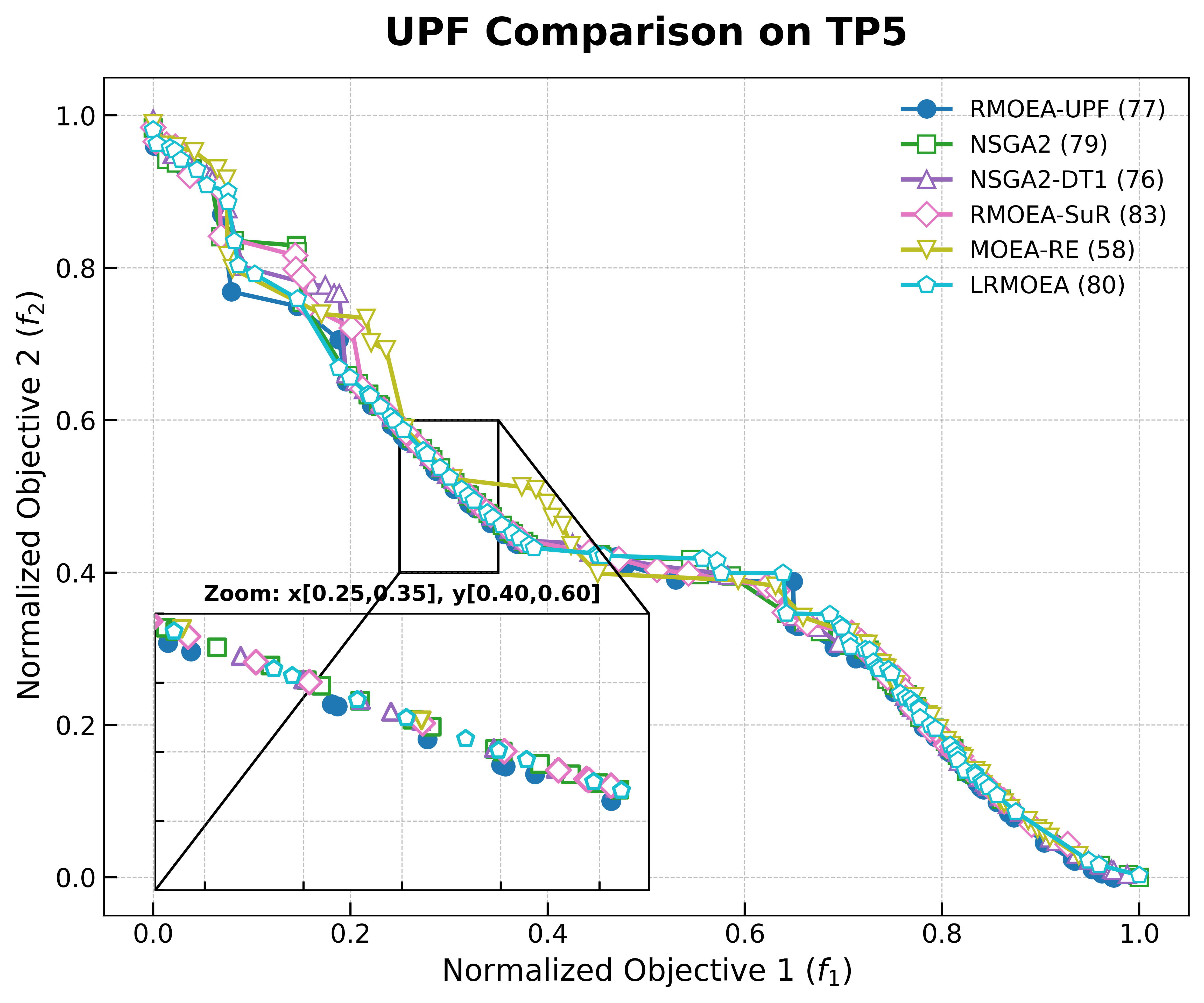}}
	\centering
	\caption{TP5 problem}
	\label{TP5}
	\vspace{-20pt}
\end{figure*}
\begin{figure*}[!h]
	\centering
	
	\subfigure[Mapping of the solution set in the objective space]{\includegraphics[width=0.47\linewidth,height=0.27\textheight]{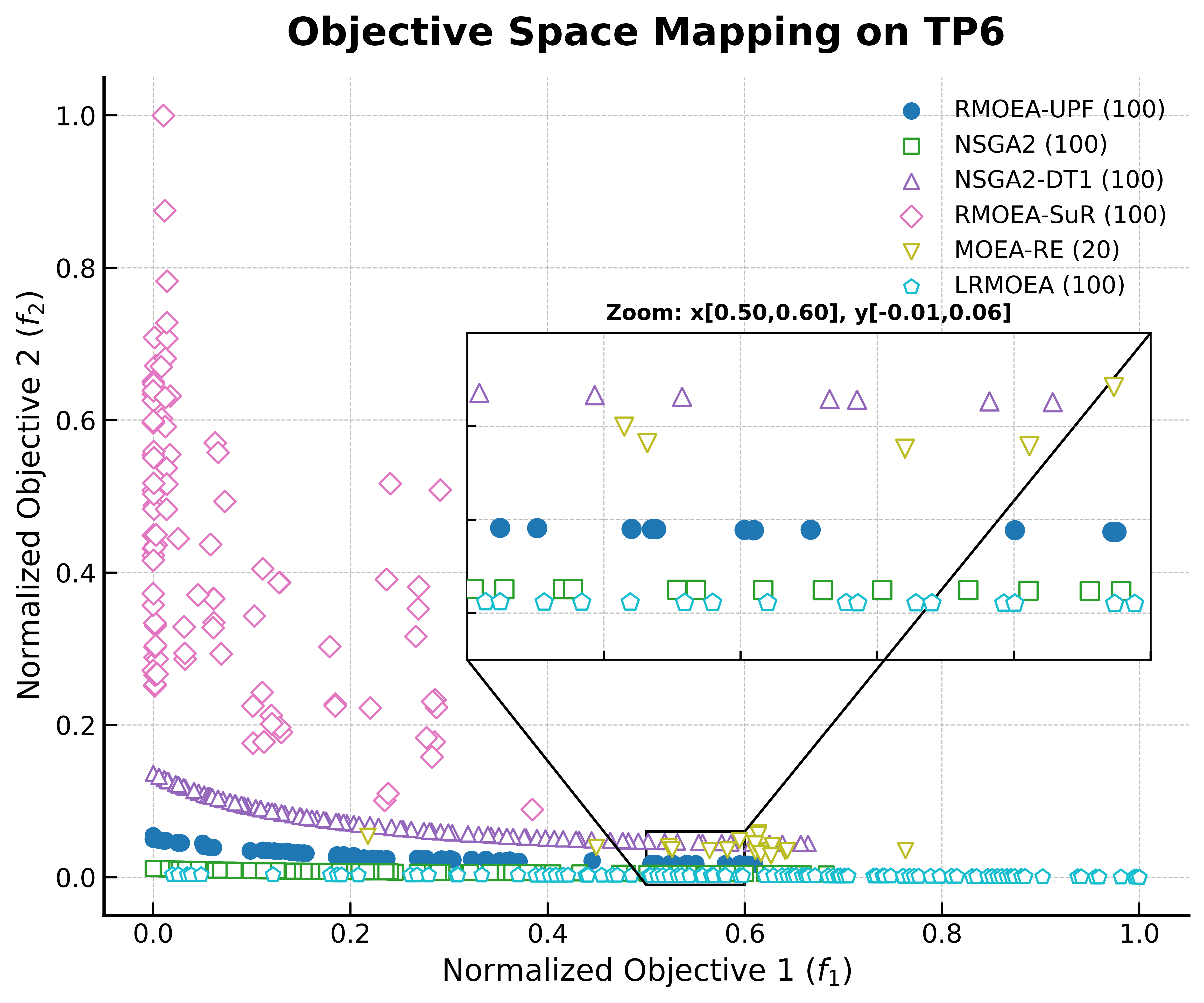}}
	\subfigure[Uncertainty-related Pareto Front]{\includegraphics[width=0.47\linewidth,height=0.27\textheight]{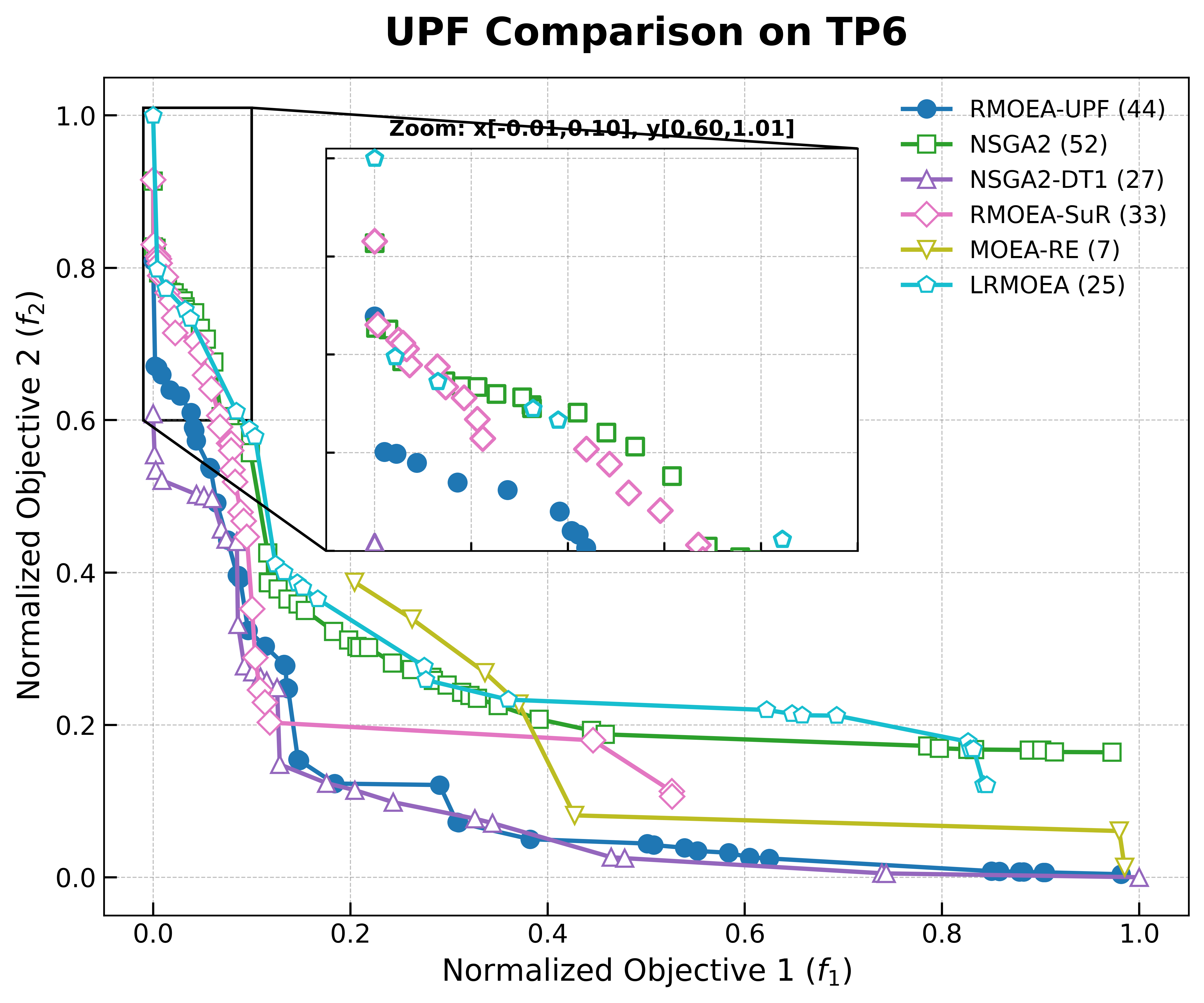}}
	\centering
	\caption{TP6 problem}
	\label{TP6}
	\vspace{-20pt}
\end{figure*}
\begin{figure*}[!h]
	\centering
	
	\subfigure[Mapping of the solution set in the objective space]{\includegraphics[width=0.47\linewidth,height=0.27\textheight]{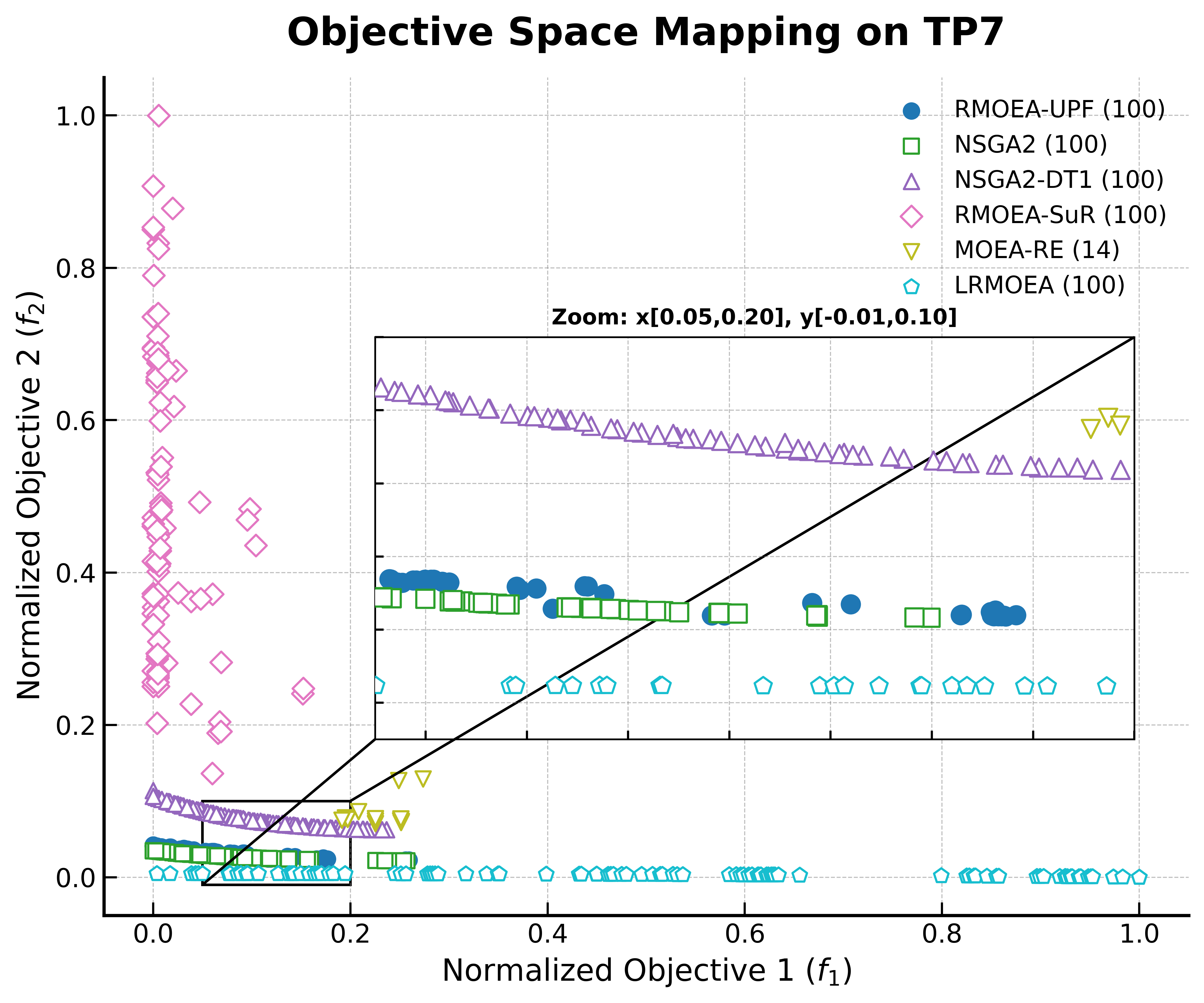}}
	\subfigure[Uncertainty-related Pareto Front]{\includegraphics[width=0.47\linewidth,height=0.27\textheight]{upf_comparison_TP7.png}}
	\centering
	\caption{TP7 problem}
	\label{TP7}
	\vspace{-20pt}
\end{figure*}
\begin{figure*}[!h]
	\centering
	
	\subfigure[Mapping of the solution set in the objective space]{\includegraphics[width=0.47\linewidth,height=0.27\textheight]{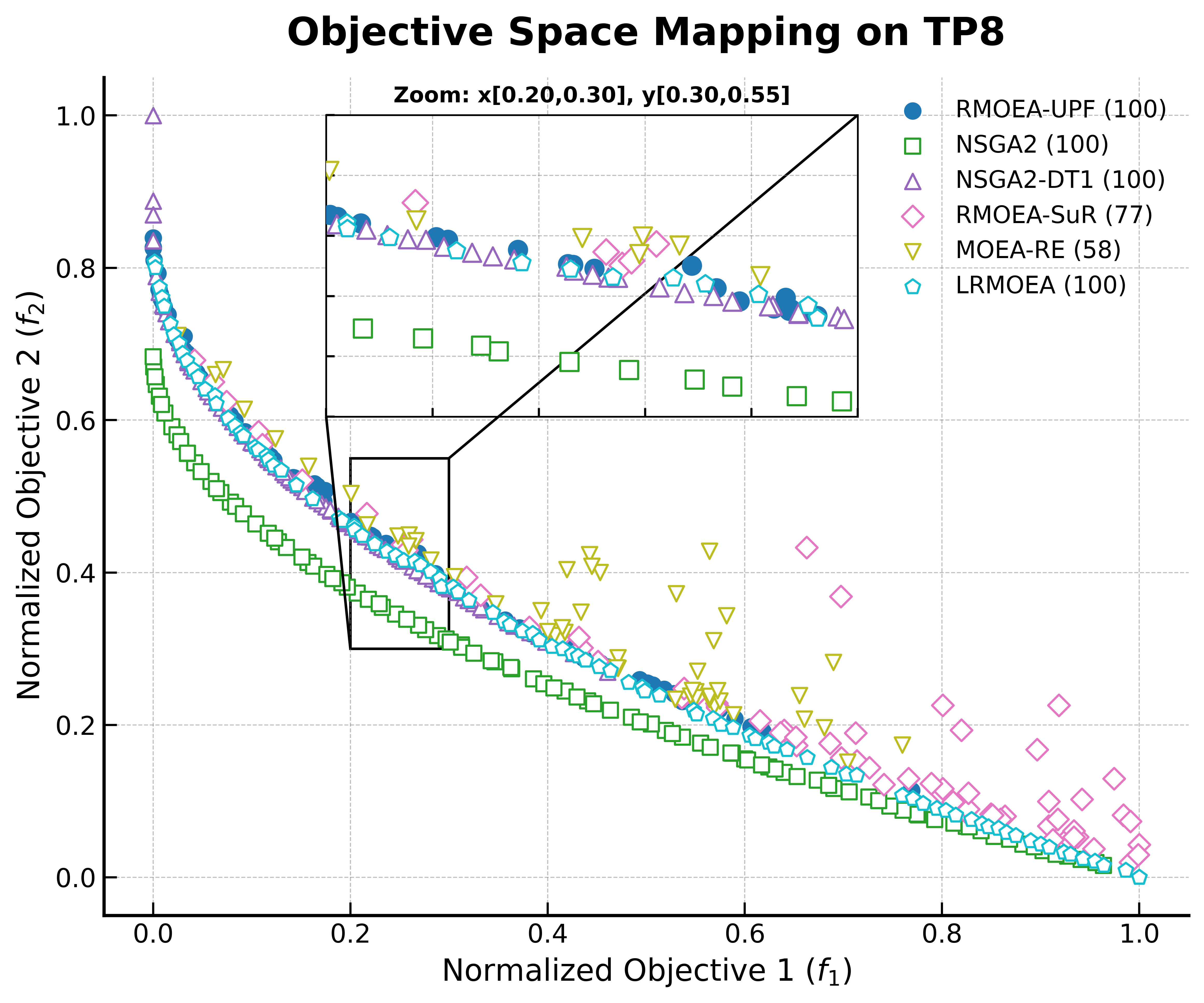}}
	\subfigure[Uncertainty-related Pareto Front]{\includegraphics[width=0.47\linewidth,height=0.27\textheight]{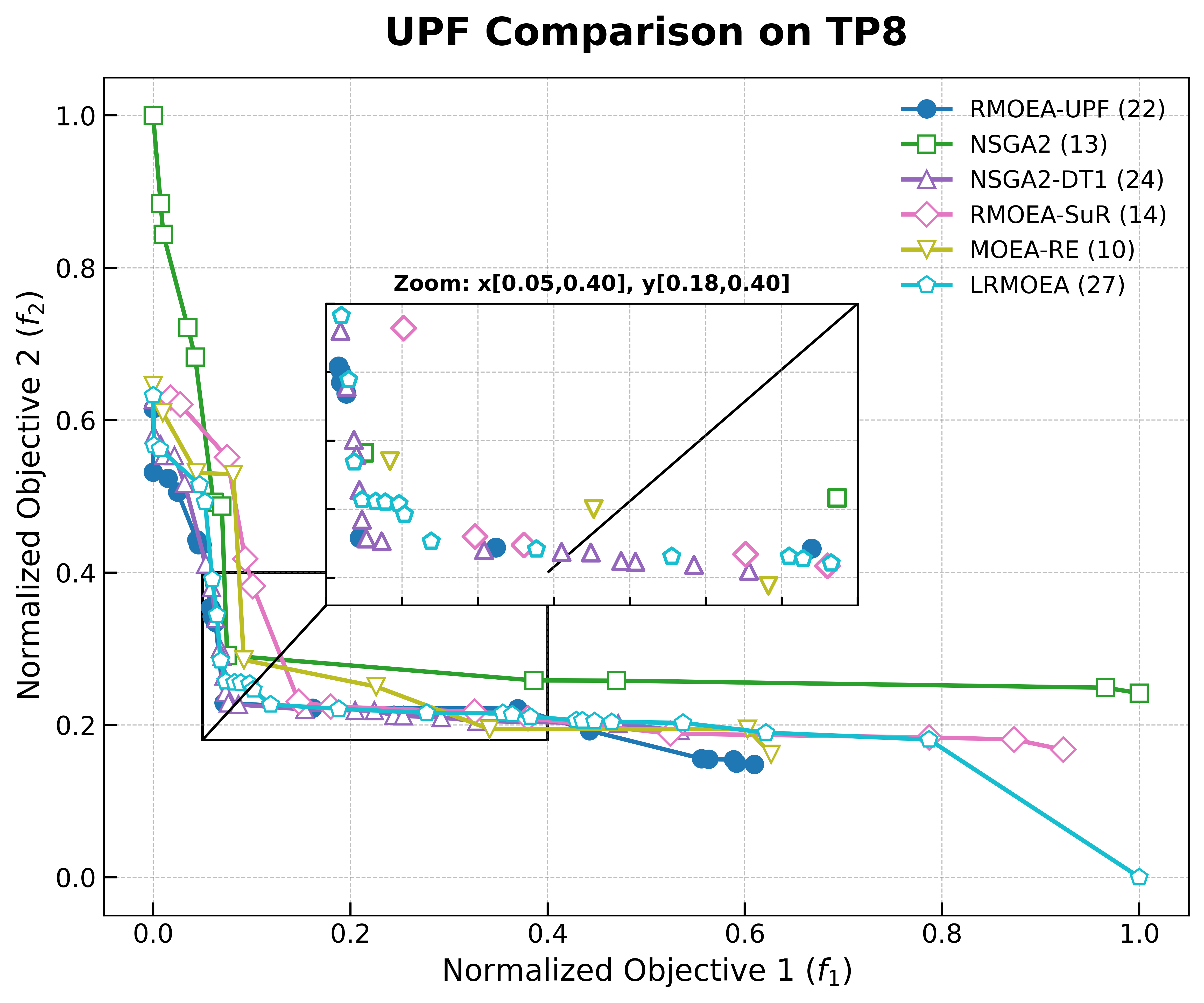}}
	\centering
	\caption{TP8 problem}
	\label{TP8}
\end{figure*}
\begin{figure*}[!h]
	\centering
	
	\subfigure[Mapping of the solution set in the objective space]{\includegraphics[width=0.47\linewidth,height=0.27\textheight]{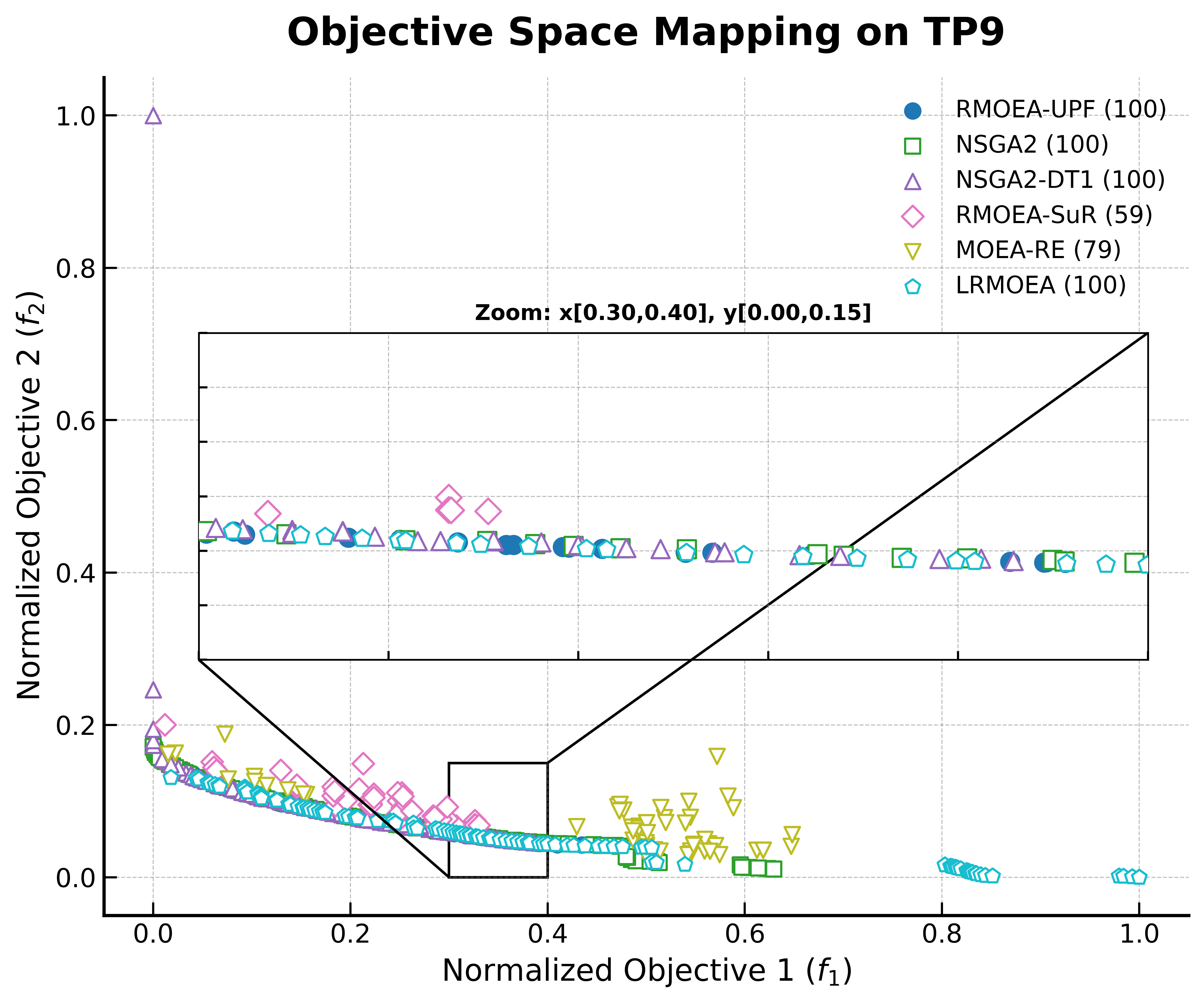}}
	\subfigure[Uncertainty-related Pareto Front]{\includegraphics[width=0.47\linewidth,height=0.27\textheight]{upf_comparison_TP9.png}}
	\centering
	\caption{TP9 problem}
	\label{TP9}
\end{figure*}

\newpage
\section{The IGD indicators of various algorithms on the TP1-TP9 problems}\label{appendix-b}
Table \ref{igd-tp1-9} presents a comparison of the IGD obtained using UPF of compared algorithms with respect to $\mathrm{UPF_{global}}$ in normalized objective space.  For test problems from TP3 to TP9, the mean IGD values of RMOEA-UPF consistently ranked among the top two. Notably, on TP3, TP5, TP7, TP8, and TP9, RMOEA-UPF achieved the best performance with the smallest IGD values. This demonstrates that RMOEA-UPF can effectively find and maintain a high-quality solution set in these complex, noise-perturbed problem environments. Additionally, on TP1, TP2, TP4, and TP6, RMOEA-UPF exhibited performance comparable to that of the comparative algorithms, including LRMOEA, MOEA-RE, RMOEA-SuR, NSGA2-DT1, and NSGA2.
\begin{table*}[!h]
	\caption{The Inverted Generational Distance (IGD) obtained using UPF of compared algorithms with respect to $\mathrm{UPF_{global}}$ in normalized objective space. }
	\begin{center}
		\renewcommand{\arraystretch}{1.5}
		\begin{tabular}{C{0.08}|C{0.135}| C{0.135}| C{0.135} |C{0.135} |C{0.135}| C{0.135}}
			
			\hline
			Problem&RMOEA-UPF&LRMOEA&MOEA-RE&RMOEA-SuR&NSGA2-DT1&NSGA2
			\\
			\hline
			TP1 &\makecell{5.926e-3 (±1.6e-4)} & \textcolor{green}{\makecell{4.770e-3 (±1.8e-4)\\($-$)}} & \makecell{7.022e-3 (±2.0e-4)\\($+$)} & \textcolor{red}{\makecell{1.414e-3 (±2.7e-4)\\($-$)}}& \textcolor{blue}{\makecell{5.610e-3 (±1.4e-4)\\($+$)}} & \makecell{6.677e-3 (±3.1e-4)\\($+$)}\\  \hline
			
			TP2 &\makecell{5.300e-3 (±1.2e-4)}& \textcolor{red}{\makecell{3.629e-3 (±8.5e-5)\\($-$)}} & \makecell{1.067e-2 (±3.5e-4)\\($+$)} &\makecell{5.183e-3 (±1.7e-4)\\($\approx$)}&  \textcolor{green}{\makecell{4.508e-3 (±1.6e-4)\\($-$)} }& \textcolor{blue}{\makecell{4.754e-3 (±1.4e-4)\\($-$)}}\\ \hline
			
			TP3 & \textcolor{red}{\makecell{4.470e-3 (±3.2e-4)}} & \makecell{5.939e-3 (±1.8e-4)\\($+$)} & \makecell{1.220e-2 (±2.9e-4)\\($+$)} & \makecell{6.018e-3 (±1.8e-4)\\($+$)} & \textcolor{green}{\makecell{5.228e-3 (±2.8e-4)\\($\approx$)}} & \textcolor{blue}{\makecell{5.913e-3 (±3.4e-4)\\($+$)}}\\\hline
			
			TP4 & \textcolor{green}{\makecell{7.081e-3 (±7.2e-4)}} & \makecell{7.947e-3 (±3.2e-4)\\($\approx$)} & \makecell{1.121e-2 (±4.9e-4)\\($+$)} & \makecell{9.314e-3 (±4.8e-4)\\($+$)} & \textcolor{blue}{\makecell{7.870e-3 (±2.0e-4)\\($\approx$)}} & \textcolor{red}{\makecell{6.864e-3 (±4.0e-4)\\($\approx$)}}\\\hline
			
			TP5 & \textcolor{red}{\makecell{5.273e-3 (±4.0e-4)}} & \textcolor{green}{\makecell{5.423e-3 (±1.9e-4)\\($\approx$)}} & \makecell{6.394e-3 (±1.6e-4)\\($+$)} & \makecell{6.625e-3 (±2.1e-4)\\($+$)} & \textcolor{blue}{\makecell{5.772e-3 (±2.3e-4)\\($\approx$)}} & \makecell{6.354e-3 (±1.2e-4)\\($+$)}\\\hline		
			
			TP6&\textcolor{green}{\makecell{3.693e-2 (±1.2e-3)}} & \makecell{1.124e-1 (±7.1e-3)\\($+$)}& \makecell{1.531e-1 (±4.4e-3)\\($+$)} & \textcolor{blue}{\makecell{8.262e-2 (±1.3e-3)\\($+$)}} & \textcolor{red}{\makecell{1.117e-2 (±2.7e-3)\\($-$)}} & \makecell{9.816e-2 (±1.1e-3)\\($+$)}\\\hline
			
			TP7 & \textcolor{red}{\makecell{9.571e-3 (±1.0e-3)}} & \makecell{4.278e-2 (±2.9e-3)\\($+$)} & \makecell{1.177e-1 (±3.8e-3)\\($+$)} & \textcolor{blue}{\makecell{2.129e-2 (±8.3e-3)\\($+$)}} & \textcolor{green}{\makecell{1.478e-2 (±3.3e-3)\\($+$)}} & \makecell{2.867e-2 (±4.4e-3)\\($+$)} \\\hline
			
			TP8 & \textcolor{red}{\makecell{1.156e-2 (±8.2e-3)}} & \textcolor{green}{\makecell{1.302e-2 (±1.6e-3))\\($\approx$)}} & \makecell{5.005e-2 (±2.6e-3)\\($+$)} & \makecell{6.895e-2 (±2.2e-3)\\($+$)} & \textcolor{blue}{\makecell{1.427e-2 (±1.2e-3)\\($\approx$)}} & \makecell{6.136e-2 (±1.3e-3)\\($+$)}\\\hline
			
			TP9 &\textcolor{red}{\makecell{3.100e-2 (±1.6e-3)}} & \makecell{5.453e-2 (±3.9e-3)\\($+$)} & \makecell{6.686e-2 (±4.8e-3)\\($+$)} & \makecell{8.844e-2 (±7.1e-3)\\($+$)} & \textcolor{blue}{\makecell{4.767e-2 (±6.9e-3)\\($+$)}} & \textcolor{green}{\makecell{4.064e-2 (±3.3e-3)\\($+$)}}\\
			
			\hline
			\multicolumn{2}{c|}{$+/\approx/-$} & \makecell{4/3/2} & \makecell{9/0/0} & \makecell{7/1/1} & \makecell{3/4/2} & \makecell{7/1/1}\\
			\hline
		\end{tabular}
		\begin{tablenotes}   
			\footnotesize             
			\item 
			We report the mean and standard deviation of IGD values over 20 independent runs. The three top-performing algorithms are highlighted using \textcolor{red}{best}, \textcolor{green}{second}, \textcolor{blue}{third}, respectively. Additionally, we use the Wilcoxon Signed Rank Test (with a significance level of 0.05) to verify whether there are significant differences between the comparative algorithms (LRMOEA, MOEA-RE, RMOEA-SuR, NSGA2-DT1, and NSGA2) and RMOEA-UPF, where ``+" indicates significantly better, ``-" indicates significantly worse, and ``$\approx$" indicates no significant difference.
		\end{tablenotes}         
		\label{igd-tp1-9}
	\end{center}
\end{table*}
\section{Parameter Sensitivity Analysis Experiment}\label{appendix-c}
\subsection{Parameter Sensitivity Analysis Experiment of Archive Capacity}
Archive capacity serves as a pivotal control parameter in the optimization workflow. It essentially quantifies the “tolerance” for accommodating solutions in the archive, where solutions are expected to exhibit a balance between convergence (proximity to the true Pareto front) and robustness (stability under perturbations). A larger archive capacity, in theory, relaxes this tolerance, allowing marginally sub-optimal solutions (in terms of immediate convergence-robustness trade-off) to enter the archive. However, a critical operational constraint arises: during archive updates, all stored solutions undergo random noise perturbation followed by real-function evaluation. This process accelerates the depletion of the maximum evaluation count. Thus, an excessively large archive capacity can truncate the effective iteration space, potentially undermining the final solution set’s convergence and robustness.

\begin{figure}[!h]
	\centering
	\includegraphics[scale=0.375]{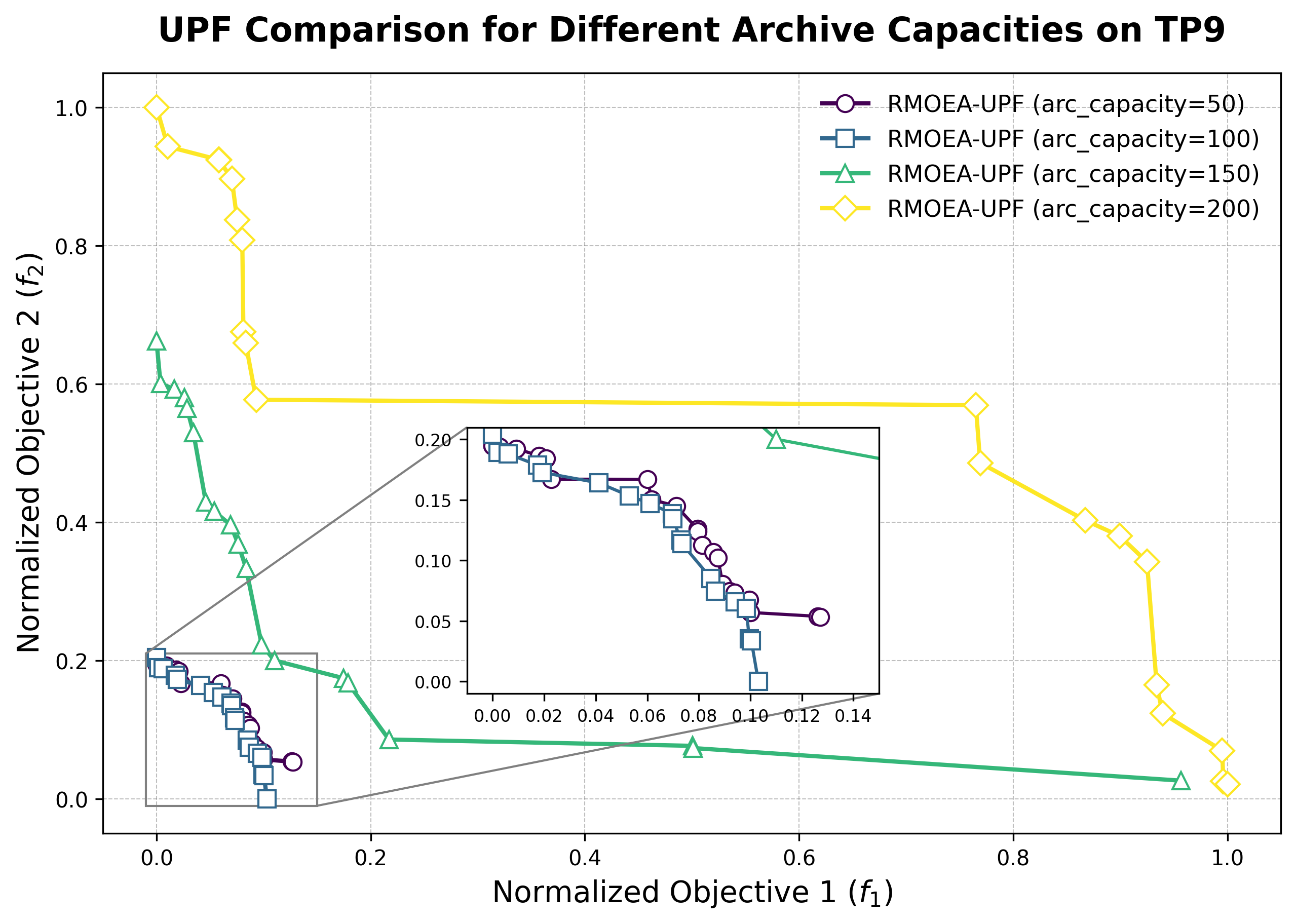}
	\caption{UPF comparison for different $arc\_capacity$}
	\label{parameter-arc-capacity-upf}
\end{figure}

We investigate the influence of archive capacity on the algorithm's performance on TP9 test problem. Here, the archive capacity (denoted as $arc\_capacity$), which dictates the maximum number of solutions with favorable convergence and robustness that can be stored during the optimization process, was set to 50, 100, 150, and 200 respectively. The evaluation count was uniformly fixed at 20000 iterations. Through this setup, we aim to dissect how different archive capacity configurations shape the final solution set's UPF within the normalized objective space. From the visualized UPF in Fig. \ref{parameter-arc-capacity-upf}, a distinct pattern emerges. For $arc\_capacity$ = 150 and $arc\_capacity$ = 200, their corresponding UPFs exhibit a clear degradation compared to the cases of $arc\_capacity$ = 50 and $arc\_capacity$ = 100. Specifically, the UPFs for larger archive capacities (150, 200) are positioned farther from the origin, indicating inferior convergence - robustness performance. When $arc\_capacity$ = 50, the UPF is slightly less optimal than that of $arc\_capacity$ = 100. This can be attributed to the overly stringent solution screening imposed by a small archive capacity. Solutions with latent potential to enhance convergence and robustness through iterative evolution may be prematurely eliminated, as they fail to meet the high-bar criteria for archive entry in the early stages. Additionally, such a restrictive setup increases the risk of the algorithm getting trapped in local optima, stifling the exploration of the global Pareto front.

The observed behavior underscores the necessity of adaptive archive capacity tuning. Since archive capacity exerts dual-edged impacts—balancing solution tolerance and iteration sufficiency—its configuration must be tailored to the specific characteristics of the problem.

\subsection{Parameter Sensitivity Analysis Experiment of Elite Offspring Size}
The elite offspring size parameter plays a dual-role in the optimization process. It directly determines the ``tolerance” for solution performance and the number of iterations. A larger $elite\_offspring\_size$ means more solutions from each generation enter the archive for subsequent updates, which is beneficial for enhancing solution diversity. However, during the archive update step, both existing solutions in the archive and current - generation elite solutions undergo real-function evaluation with noise perturbation. Thus, a larger $elite\_offspring\_size$ can reduce the effective number of iterations. Hence, like the archive capacity, $elite\_offspring\_size$ needs to be adaptively set according to specific problems.

In the exploration of multi - objective optimization for the TP9 test problem, we focus on the impact of the $elite\_offspring\_size$ parameter. This parameter, representing the number of solutions selected from offspring to enter the archive in each generation, was set to 10, 20, 30, 40, and 50 respectively. Through setting different $elite\_offspring\_size$ values, we aim to reveal how this parameter shapes the Uniformly - distributed Pareto Front (UPF) of the final solution set in the normalized objective space. 

\begin{figure}[!h]
	\centering
	\includegraphics[scale=0.375]{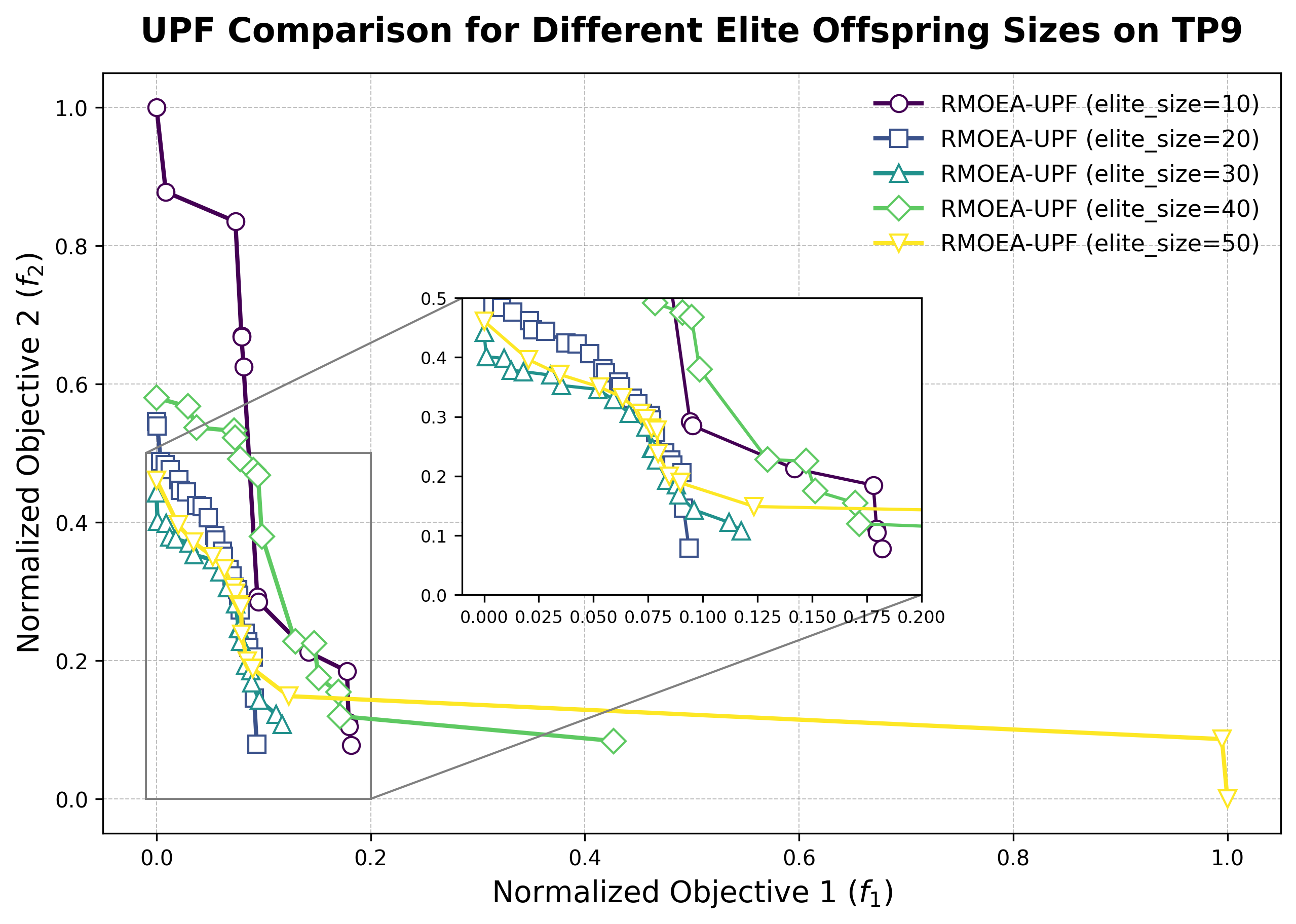}
	\caption{UPF comparison for different $elite\_offspring\_size$}
	\label{parameter-elite-size-upf}
\end{figure}

From the visualized UPF in Fig. \ref{parameter-elite-size-upf}, a clear pattern emerges. When $elite\_offspring\_size$ = 30, the corresponding UPF is the closest to the origin, indicating the best convergence and robustness performance. In contrast, when $elite\_offspring\_size$ = 10, the UPF is the farthest from the origin, representing the worst performance. For $elite\_offspring\_size$ = 10, the poor performance may result from the insufficient number of elite solutions entering the archive per generation. With a lack of ``fresh blood” (new elite solutions), the iteration process mainly relies on old solutions already fixed in the archive when generating parent solutions. This situation easily leads the algorithm into the trap of local optimality, as the exploration of the solution space is severely limited.
\end{document}